\documentclass[journal]{IEEEtran}

\usepackage{cite}
\usepackage{graphicx}
\usepackage{mathrsfs}
\usepackage{amssymb}
\usepackage{amsmath}
\usepackage{subeqnarray}
\usepackage{cases}
\usepackage{booktabs}
\usepackage{amsthm}
\usepackage{caption}
\usepackage{subfigure}
\usepackage{algorithm}
\usepackage{algorithmic}
\usepackage{url}
\usepackage{amsmath,amssymb}
\usepackage{graphicx}
\usepackage{url}
\usepackage{booktabs}
\usepackage{multirow}
\usepackage{subfigure}
\usepackage{authblk}
\usepackage{amsthm}
\usepackage{bm}
\usepackage{algorithm}
\usepackage{stfloats}
\usepackage{array}
\usepackage{eqparbox}
\usepackage{bm}
\usepackage{lipsum}
\allowdisplaybreaks[4]

\makeatletter
\renewcommand{\maketag@@@}[1]{\hbox{\m@th\normalsize\normalfont#1}}%
\makeatother

\ifCLASSINFOpdf

\else

\fi

\begin{document}
\title{NOMA-Enabled Multi-Beam Satellite Systems: Joint Optimization to Overcome Offered-Requested Data Mismatches}

\author{Anyue~Wang,
~\IEEEmembership{Student member,~IEEE,}
        Lei~Lei,
~\IEEEmembership{Member,~IEEE,}
        Eva~Lagunas,
~\IEEEmembership{Senior Member,~IEEE,}
        Ana~I.~P\'{e}rez-Neira,
~\IEEEmembership{Fellow,~IEEE,}
        Symeon~Chatzinotas,
~\IEEEmembership{Senior Member,~IEEE,}
        and~Bj\"{o}rn~Ottersten,
~\IEEEmembership{Fellow,~IEEE}% <-this % stops a space
\thanks{A. Wang, L. Lei, E. Lagunas, S. Chatzinotas, and B. Ottersten are with Interdisciplinary Center for Security, Reliability and Trust, University of Luxembourg, 1855 Luxembourg (email: anyue.wang; lei.lei; eva.lagunas; symeon.chatzinotas; bjorn.ottersten@uni.lu).}% <-this % stops a space
\thanks{A. I. P\'{e}rez-Neira is with the Centre Tecnol\`{o}gic de Telecomunicacions de Catalunya (CTTC/CERCA), 08860 Castelldefels, Spain, and Universitat Polit\`{e}cnica de Catalunya (UPC), 08034 Barcelona, Spain (email: ana.perez@cttc.es).}
\thanks{A part of this work was presented at IEEE International Symposium on Personal, Indoor and Mobile Radio Communications (PIMRC), 2019. The work has been funded by the FNR CORE project ROSETTA (11632107) and FlexSAT (C19/IS/13696663).}}% <-this % stops a space
%\thanks{}}

\maketitle

\begin{abstract}
Non-Orthogonal Multiple Access (NOMA) has potentials to improve the performance of multi-beam satellite systems.
The performance optimization in satellite-NOMA systems can be different from that in terrestrial-NOMA systems, e.g., considering distinctive channel models, performance metrics, power constraints, and limited flexibility in resource management. 
In this paper, we adopt a metric, Offered Capacity to requested Traffic Ratio (OCTR), to measure the requested-offered data (or rate) mismatch in multi-beam satellite systems.
In the considered system, NOMA is applied to mitigate intra-beam interference while precoding is implemented to reduce inter-beam interference.
We jointly optimize power, decoding orders, and terminal-timeslot assignment to improve the max-min fairness of OCTR.
The problem is inherently difficult due to the presence of combinatorial and non-convex aspects.
We first fix the terminal-timeslot assignment, and develop an optimal fast-convergence algorithmic framework based on Perron-Frobenius theory (PF) for the remaining joint power-allocation and decoding-order optimization problem.
Under this framework, we propose a heuristic algorithm for the original problem, which iteratively updates the terminal-timeslot assignment and improves the overall OCTR performance.
Numerical results verify that max-min OCTR is a suitable metric to address the mismatch issue, and is able to improve the fairness among terminals.
In average, the proposed algorithm improves the max-min OCTR by 40.2\% over Orthogonal Multiple Access (OMA).
\end{abstract}


\IEEEpeerreviewmaketitle

\section{Introduction}
\IEEEPARstart{A} multi-beam satellite system provides wireless service to wide-range areas.
On the one hand, traffic distribution is typically asymmetric among beams \cite{perez2019signal}.
On the other hand, satellite capacity is restricted by practical aspects, e.g., payload design, limited flexibility in resource management, and tended to be fixed before launch \cite{kodheli2020satellite}.
The asymmetric traffic and the predesigned capacity could result in mismatches between requested traffic and offered capacity \cite{simulator}, i.e., hot beams with unmet traffic demand or cold beams with unused capacity \cite{cocco2017radio}.
Both cases are undesirable for satellite operators, which motivates the investigation of flexible resource allocation to reduce the mismatches for future multi-beam satellite systems. 

In terrestrial systems, Non-Orthogonal Multiple Access (NOMA) has demonstrated its superiority, e.g., in throughput, energy, fairness, etc., \cite{lei2019learning, islam2017power}, over Orthogonal Multiple Access (OMA). 
By performing superposition coding at the transmitter side, more than one terminal's signal can be superimposed with different levels of transmit power and broadcast to co-channel allocated terminals.
At the receiver side, Successive Interference Cancellation (SIC) is performed.
In this way, NOMA is capable of alleviating co-channel interference, accommodating more terminals, and improving spectrum efficiency \cite{islam2017power}.
In satellite systems, NOMA has attracted early studies, e.g., \cite{okamoto2016application, caus2016noma, perez2019non, ugolini2019capacity, yan2019application, zhu2017non, lin2019joint, beigi2018interference, yan2018outage, liu2019qos}.
The authors in \cite{okamoto2016application, caus2016noma, perez2019non, ugolini2019capacity, yan2019application} analyzed the applicability of integrating NOMA to satellite systems.
In \cite{okamoto2016application}, NOMA was applied in satellite-terrestrial integrated systems to improve  capacity and fairness.
NOMA was considered in multi-beam satellite systems in \cite{caus2016noma} and \cite{perez2019non}, where precoding, power allocation, and user grouping schemes were studied to maximize the capacity.
The authors in \cite{ugolini2019capacity} investigated capacity improvement by the technique of multi-user detection for non-orthogonal transmission in multi-beam satellite systems. 
In \cite{yan2019application}, the authors provided an overview for applying power-domain NOMA to satellite networks.
%In \cite{perez2019non}, the authors discussed the application of NOMA coexisting with different frequency-reuse schemes and addressed one of the open issues, i.e., NOMA for multicast scenarios.

In the literature, resource optimization for NOMA-enabled multi-beam satellite systems is studied to a limited extent.
For instance, in \cite{zhu2017non} and \cite{lin2019joint}, user pairing, precoding design, and power allocation were investigated for NOMA-based satellite-terrestrial integrated systems, where the satellite is functioned as a supplemental component.
In both works, NOMA was implemented in the terrestrial part whereas OMA was adopted in the satellite part.
The authors in \cite{beigi2018interference} proposed a NOMA scheme at the beam level, via the cooperation of neighboring beams to improve the capacity. 
For mathematical analysis, the authors in \cite{yan2018outage} studied the outage performance of NOMA-based satellite-terrestrial integrated systems.
The above works commonly adopted general metrics, e.g., capacity, fairness, and outage probability.
Nevertheless, the metrics capturing the matches between requested traffic and offered capacity, have not been fully discussed yet.
The authors in \cite{liu2019qos} studied power optimization for NOMA-based multi-beam satellite systems, with adopting a predefined and fixed decoding order, thus simplifying the power allocation.
In practical scenarios, decoding orders may change when other beams' power is adjusted \cite{you2018resource}.
Therefore, it is important to optimize decoding orders for multi-beam satellite-NOMA systems since an inappropriate decoding order can result in unsuccessful SIC and thus performance degradation.
In this paper, we consider a full frequency reuse system, where inter-beam interference is mitigated via precoding while NOMA is applied to reduce intra-beam interference within a beam.

In general, resource allocation schemes for terrestrial multi-antenna NOMA systems may not be directly applied to multi-beam satellite systems \cite{kodheli2020satellite, perez2019non}.
For instance, terminals with highly correlated channels and large channel gain difference are favorable to be grouped to mitigate inter-beam and intra-beam interference by precoding and NOMA, respectively \cite{liu2017joint, chinnadurai2017novel,ali2016non}.
Such desired terminal groups or pairs can be observed in terrestrial-NOMA systems but might not be easily obtained in satellite scenarios.
In addition, channel models, payload design, and on-board limitations could render resource optimization in satellite-NOMA systems more challenging than terrestrial-NOMA systems \cite{aravanis2015power}.

In our previous work \cite{wang2019fairness}, we focus on power allocation in multi-beam satellite-NOMA systems, and develop a heuristic power-tune algorithm, without convergence guarantee, to improve the performance of Offered Capacity to requested Traffic Ratio (OCTR).
In comparison, the major improvement of this paper is that, first, we jointly optimize power, decoding orders, and terminal-timeslot assignment, which brings more performance gain of OCTR but is much more challenging than the addressed problem in \cite{wang2019fairness}.
Second, we augment the power-tune solution by deriving theoretical results such that fast convergence can be guaranteed.
Third, we provide a complete algorithmic solution for the considered joint optimization problem, instead of only power solution in \cite{wang2019fairness}.
%In this paper, we apply NOMA to address a practical issue in satellite scenarios, i.e., the mismatches between offered capacity and requested traffic.
%We focus on resource optimization by taking into account practical limitations in satellite systems.
%The objective of the formulated optimization problem is a max-min fairness design applied to Offered Capacity to requested Traffic Ratio (OCTR).
%In our previous work \cite{wang2019fairness}, we have discussed power optimization to improve the max-min fairness of OCTR for multi-beam satellite systems.
%In the paper, we will further study this metric and investigate the joint optimization of not only power allocation, but also decoding orders and terminal-timeslot assignment.
The main contributions are summarized as follows:
\begin{itemize}
\item We formulate a max-min resource allocation problem to jointly optimize power allocation, decoding orders, and terminal-timeslot assignment, such that the lowest OCTR among terminals can be maximized.
The problem falls into the domain of combinatorial non-convex programming.
\item Power optimization in NOMA-based multi-beam/cell systems typically encounters the issues of undetermined optimal decoding order, and undetermined rate-function expressions.
In this work, based on the derived theoretical analysis, we circumvent these difficulties and provide a simple approach to tackle the above issues.
\item By fixing the terminal-timeslot assignment, we propose a Perron-Frobenius theory (PF) based approach to solve the remaining problem, i.e., Jointly Optimizing Power allocation and Decoding orders (JOPD).
The approach is proven with guaranteed fast convergence to the optimum.
The fixed terminal-timeslot assignment is determined by grouping the terminals with Maximum Channel Correlation (MaxCC).
\item Under the framework of JOPD, we develop a heuristic algorithm to Jointly Optimizing Power allocation, Decoding orders, and Terminal-timeslot scheduling (JOPDT), which iteratively updates terminal-timeslot assignment, precoding vectors, and improves the overall OCTR performance.
\item The numerical results, firstly, verify the fast convergence of JOPD.
Secondly, we show the OCTR performance gain of NOMA over OMA in two NOMA-based schemes, i.e., JOPD+MaxCC (with lower complexity) and JOPDT (with higher complexity).
Lastly, the results show that the max-min OCTR can be an appropriate metric to address the mismatch issue and enhance terminals' fairness.
\end{itemize}

The remainder of the paper is organized as follows: 
Section II introduces the system model of NOMA-enabled multi-beam satellite systems.
The max-min optimization problem is formulated and the challenges of the problem solving are discussed in Section III.
We propose a PF-based algorithmic framework, JOPD, to solve the problem with the fixed terminal-timeslot scheduling in Section IV, where the convergence and optimality of JOPD are analyzed.
Besides, we discuss the strategies of assigning each timeslot to terminals.
In Section V, the heuristic algorithm JOPDT is put forward to solve the original problem.
The simulation settings are displayed and the numerical results are analyzed in Section VI.
Section VII concludes the paper.

The notations in this paper are as follows:
The operators $[\cdot]^T$ and $[\cdot]^H$ denote the transpose and conjugate transpose operator, respectively.
$\vert\cdot\vert$ represents the cardinality of a set or the absolute value.
$\Vert\cdot\Vert$ denotes the Euclidean norm of a vector.
$[\cdot]_{i,j}$ represents the element in the $i$-th row and the $j$-th column of a matrix.

\section{System Model}
\subsection{A Multi-Beam Satellite System}
We consider forward-link transmission in a multi-beam satellite system, where a Geostationary Earth Orbit (GEO) satellite is equipped with an array-fed reflector antenna to generate $B$ spot beams.
We denote $\mathcal{B}=\lbrace 1,\dots,B\rbrace$ as the set of the beams.
One feed per beam is implemented in the system and the index of a feed is assumed to be consistent with that of the beam it serves.
Let ${\mathcal{U}}_b$ be the set of all the terminals located within the service area of the $b$-th beam.
Time Division Multiple Access (TDMA) mode is applied in the system.
We focus on resource allocation during a scheduling period consisting of $C$ timeslots.
Let $\mathcal{C}=\lbrace 1,\dots,C\rbrace$ be the set of the timeslots.
For each scheduling period, $K_b$ terminals are selected for transmission.
Denote $\mathcal{K}_b=\lbrace 1,\dots,K_b\rbrace$ as the set of the selected terminals in beam $b$, where $\mathcal{K}_b\subseteq \mathcal{U}_b$.
The satellite provides Fixed Satellite Services (FSS) to ground terminals, where the channel gains vary over scheduling periods but keep static during a scheduling period.
Define $\mathbf{h}_{bk}=[h_{bk}^1,\dots,h_{bk}^i,\dots,h_{bk}^B]^T\in\mathbb{C}^{B\times 1}$ as the channel vector of the $k$-th terminal in beam $b$ at timeslot $c$.
The $i$-th element of the vector, $h_{bk}^{i}$, denotes the channel coefficient from the $i$-th feed to the $k$-th terminal in beam $b$, where $i\in\mathcal{B}$.
The channel coefficient can be expressed as 
$h_{bk}^{i}=G^{\textrm{Sat}}_{ibk}L_{bk}G^{\textrm{Rx}}_{bk}$, where $G^{\textrm{Sat}}_{ibk}$ is the transmit antenna gain corresponding to the off-axis angle between the beam center and the terminal.
$L_{bk}$ is the free-space propagation loss from the satellite to the terminal.  
$G^{\textrm{Rx}}_{bk}$ is the receiver antenna gain. 
By introducing NOMA and precoding to mitigate interference, 1-color frequency-reuse pattern is adopted in this work.
In terms of payload, the on-board payload is equipped with the module of Multi-Port Amplifier (MPA)  such that power can be flexibly distributed across different beams.

\subsection{Precoding and NOMA}
To alleviate inter-beam interference, we adopt a linear precoding scheme, Minimum Mean Square Error (MMSE), which is considered with high efficiency and low computational complexity \cite{caus2016noma}.
Denote $\mathbf{w}_{bc}=[w_{bc}^1,\dots,w_{bc}^i,\dots,w_{bc}^B]^T\in\mathbb{C}^{B\times 1}$ as the precoding vector for the $b$-th beam at timeslot $c$.
The $i$-th element of the vector, $w_{bc}^i$, represents the precoding coefficient of the $i$-th feed for the $b$-th beam, where $i\in\mathcal{B}$.
The received signal can be expressed as:
\begin{align}
y_{bkc}=&\underbrace{\mathbf{h}_{bk}^H\mathbf{w}_{bc}\sqrt{p_{bkc}}s_{bkc}}_{\textrm{desired signal}}+\underbrace{\sum_{l\in\mathcal{K}_b\setminus \lbrace k\rbrace}\mathbf{h}_{bk}^H\mathbf{w}_{bc}\sqrt{p_{blc}}s_{blc}}_{\textrm{intra-beam interference}}\notag\\
&+\underbrace{\sum_{b^{'}\in\mathcal{B}\setminus \lbrace b\rbrace}\sum_{j\in\mathcal{K}_{b^{'}}}\mathbf{h}_{bk}^H\mathbf{w}_{b^{'}c}\sqrt{p_{b^{'}jc}}s_{b^{'}jc}}_{\textrm{inter-beam interference}}+\underbrace{n_{bkc}}_{\textrm{noise}},
\label{equ:rs}
\end{align}
where $s_{bkc}$, $p_{bkc}$, and $n_{bkc}\sim\mathcal{CN}(0,\sigma^2)$ are the signal with unit power, power scaling factor, and the complex circular symmetric independent identically distributed Additive White Gaussian Noise (AWGN) with zero mean and variance $\sigma^2$, respectively.
%Due to the limited capability of the satellite payload, the transmit power of a feed is identical across timeslots \cite{destounis2011dynamic}.
The transmit power of the $b$-th beam (or feed) is $\rho_{bc}\sum_{k\in\mathcal{K}_b}p_{bkc}$, $\forall c\in\mathcal{C}$, where $\rho_{bc}=[\sum_{i\in\mathcal{B}}\mathbf{w}_{ic}\mathbf{w}_{ic}^H]_{b,b}$ denotes the power radiated by the $b$-th feed for precoding \cite{christopoulos2015multicast}.

To implement MMSE, we construct $\mathbf{H} \in \mathbb{C}^{B\times B}$ as the channel matrix, where the $b$-th row represents the channel vector of the terminal with $\max_{k\in\mathcal{K}_b}\Vert\mathbf{h}_{bk}\Vert$ \cite{ali2016non}.
The precoding matrix reads,
\begin{equation}
\mathbf{W}=\beta\mathbf{H}^H(\mathbf{H} \mathbf{H}^H+\sigma^2\mathbf{I}_{B})^{-1},
\end{equation}
where $\mathbf{I}_B$ is the identity matrix with the dimension $B$ by $B$. 
$\beta$ is a scaling factor to normalize the precoding matrix as $[\mathbf{W}\mathbf{W}^H]_{b,b}\leq1$, $\forall b\in\mathcal{B}$.
The scaling factor can be determined as $\beta^2=\frac{1}{\max\lbrace\textrm{diag}((\mathbf{H}^H\mathbf{H})^{-1})\rbrace}$.
Note that the regularization factor before $\mathbf{I}$ is fixed to $\sigma^2$ in this paper.

Within a beam, NOMA is applied to mitigate intra-beam interference among terminals.
We use $\phi_{bklc}\in\lbrace 0,1\rbrace$ to indicate decoding orders, where $k\neq l$.
If terminal $k$ is able to decode and remove the signals of $l$ before decoding its own signals, $\phi_{bklc}=0$, otherwise, $\phi_{bklc}=1$.
The Signal-to-Interference-plus-Noise Ratio (SINR) $\gamma_{bkc}$ is expressed as in \eqref{sinr}.
\begin{equation}
\gamma_{bkc}=\frac{\vert\mathbf{h}_{bk}^H\mathbf{w}_{bc}\vert^2p_{bkc}}{\hspace{-0.4cm}\sum\limits_{l\in\mathcal{K}_b\setminus \lbrace k\rbrace}\hspace{-0.4cm}\phi_{bklc}\vert\mathbf{h}_{bk}^H\mathbf{w}_{bc}\vert^2p_{blc}+\hspace{-0.2cm}\sum\limits_{b^{'}\in\mathcal{B}\setminus \lbrace b \rbrace}\hspace{-0.4cm}\vert\mathbf{h}_{bk}^H\mathbf{w}_{b^{'}c}\vert^2 \hspace{-0.25cm}\sum\limits_{j\in\mathcal{K}_{b^{'}}}\hspace{-0.2cm}p_{b^{'}jc}\hspace{-0.15cm}+\sigma^2}.
\label{sinr}
\end{equation}

According to a widely-adopted approach for determining decoding orders \cite{islam2017power, you2018resource, liu2017joint, wang2019fairness, lei2019load}, the SIC decoding order is the descending order of the ratio between channel gain and inter-beam interference plus noise.
The ratio of terminal $k$ in beam $b$ at timeslot $c$ is denoted by,
\begin{equation}
g_{bkc}=\dfrac{\vert\mathbf{h}_{bk}^H\mathbf{w}_{bc}\vert^2}{\sum\limits_{b^{'}\in\mathcal{B}\setminus \lbrace b \rbrace}\vert\mathbf{h}_{bk}^H\mathbf{w}_{b^{'}c}\vert^2 \sum\limits_{j\in\mathcal{K}_{b^{'}}}p_{b^{'}jc}+\sigma^2}.
\label{equ:g}
\end{equation}
To ease the presentation, we assume the decoding order is consistent with the terminal index, i.e., $g_{b1c}\geq g_{b2c}\geq\dots\geq g_{bK_bc}$, unless otherwise stated.

The throughput of terminal $k$ in beam $b$ at timeslot $c$ is,
\begin{equation}
R_{bkc}=B_W\log(1+\gamma_{bkc}),
\label{equ:r}
\end{equation}
where $B_W$ is the bandwidth that is occupied.
Hence the offered capacity of that terminal is derived as,
\begin{equation}
R_{bk}=\sum_{c\in\mathcal{C}}R_{bkc}.
\end{equation}

\section{Problem Formulation}
We formulate a max-min fairness problem to improve the OCTR performance by power, decoding-order, and terminal-timeslot optimization.
We define the variables and formulate the max-min fairness problem $\mathcal{P}_0$ as follows:
\begin{subequations}
\begin{align}
p_{bkc}&\geq 0,\,\,\,\,\textrm{allocated power for terminal }k\textrm{ in beam }b\notag\\
&\,\,\,\,\,\,\,\,\,\,\,\,\,\,\,\textrm{at timeslot }c,\notag\\
\phi_{bklc}&=\begin{cases}
0, \,\,\,\,\textrm{in beam }b\textrm{, terminal }k\textrm{ is able to decode the}\\\,\,\,\,\,\,\,\,\textrm{ signals of }l\textrm{ at timeslot }c,\\ 
1, \,\,\,\,\textrm{otherwise},
\end{cases}\notag\\
\alpha_{bkc}&=\begin{cases}
1, \,\,\,\,\textrm{terminal }k\textrm{ in beam }b\textrm{ is scheduled to time-}\\\,\,\,\,\,\,\,\,\,\,\textrm{slot }c,\\ 
0, \,\,\,\,\textrm{otherwise},
\end{cases}\notag\\
    \mathcal{P}_0:&\max_{p_{bkc},\phi_{bklc},\alpha_{bkc}}\,\,\,\,\min_{b\in\mathcal{B},k\in\mathcal{K}_b}\,\,\,\,  \frac{R_{bk}}{D_{bk}} \label{OPT1}\\
    \,\,\,\,\mbox{s.t.}\,\,\,\,
    &\sum_{b\in\mathcal{B}}\rho_{bc}\sum_{k\in\mathcal{K}_b}p_{bkc}\leq P_{tot},\forall c \in\mathcal{C},\label{con:pt1}\\
    &\rho_{bc}\sum_{k\in\mathcal{K}_b}p_{bkc}\leq P_{b,\max},\forall b\in\mathcal{B}, \forall c\in\mathcal{C}, \label{con:pbm1}\\
    &\rho_{bc}\sum_{k\in\mathcal{K}_b}p_{bkc}= \rho_{bc^{'}}\sum_{k\in\mathcal{K}_b}p_{bkc^{'}},\notag\\
    &\,\,\,\,\,\,\,\,\,\,\,\,\,\,\,\,\,\,\,\,\,\,\,\,\,\,\,\,\,\,\,\,\,\,\,\,\,\,\,\,\,\,\,\,\,\,\,\,\,\,\,\,\,\,\,\,\forall b\in\mathcal{B},\forall c,c^{'}\in\mathcal{C}, c\neq c^{'},\label{con:pc1}\\
    &\sum_{k\in\mathcal{K}_b}\alpha_{bkc}\leq \bar{K},\forall b\in\mathcal{B}, \forall c\in\mathcal{C}, \label{con:al1}\\
    &\sum_{c\in\mathcal{C}}\alpha_{bkc}=1, \forall b\in\mathcal{B}, \forall k\in\mathcal{K}_b, \label{con:ak1}\\
    &p_{bkc}\leq \hat{P}\alpha_{bkc},\forall b\in\mathcal{B},\forall c\in\mathcal{C},\forall k\in\mathcal{K}_b,\label{con:pa1}\\
    &g_{blc}-g_{bkc}\leq A\phi_{bklc}, \notag\\
    &\,\,\,\,\,\,\,\,\,\,\,\,\,\,\,\,\,\,\,\,\,\,\,\,\,\,\,\,\forall b\in\mathcal{B}, \forall c\in\mathcal{C}, \forall k, l\in\mathcal{K}_b,k\neq l, \label{con:phig1}\\
    &\phi_{bklc}+\phi_{blkc}=1, \notag\\
    &\,\,\,\,\,\,\,\,\,\,\,\,\,\,\,\,\,\,\,\,\,\,\,\,\,\,\,\,\forall b\in\mathcal{B}, \forall c\in\mathcal{C}, \forall k, l\in\mathcal{K}_b, k\neq l. \label{con:phi1}
\end{align}
\end{subequations}
In the objective, we focus on the OCTR improvement and fairness enhancement at the terminal level \cite{9014019}.
The OCTR metric for terminal $k$ in beam $b$ is defined as 
$\frac{R_{bk}}{D_{bk}}$, where $R_{bk}$ and $D_{bk}$ are the offered capacity and requested traffic demand, respectively.
The optimization task is to maximize the worst OCTR among terminals in $\mathcal{K}_b$, such that the mismatch and the fairness issues can be addressed. 
%Note that we consider all the terminals in $\mathcal{K}_b$ in the optimization.
In \eqref{con:pt1}, the total power is less than a budget $P_{tot}$, due to the limited on-board power supply.
Constraints \eqref{con:pbm1} state that the allocated power for each beam should be restricted by the power constraint, $P_{b,\max}$.
Constraints \eqref{con:pc1} denote that, the power allocated to each beam is identical across timeslots, considering the practical issues in waveform design, dynamic range of the signal, and non-linearities of the amplifier \cite{kodheli2020satellite, aravanis2015power, destounis2011dynamic}.
For each beam, the number of terminals simultaneously accessing the same timeslot is no more than $\bar{K}$ in \eqref{con:al1}.
In \eqref{con:ak1}, each terminal is limited to be scheduled once during a scheduling period to avoid imbalanced timeslot assignment among terminals, which is important for serving a large number of terminals.
Constraints \eqref{con:pa1} connect two sets of variables, $p_{bkc}$ and $\alpha_{bkc}$, where $\hat{P}$ is no smaller than the maximal $p_{bkc}$, e.g., $\hat{P}=P_{{tot}}$.
If $\alpha_{bkc}=0$, $p_{bkc}$ is zero.
If $\alpha_{bkc}=1$, $\hat{P}\geq p_{bkc}>0$ since the option $\alpha_{bkc}=1$ and $p_{bkc}=0$ is clearly not optimal, thus will be excluded from the optimum.
Constraints \eqref{con:phig1} and \eqref{con:phi1} confine variables $\phi_{bklc}$ to perform SIC by the descending order defined in \eqref{equ:g}, where $A$ is no smaller than the maximum value of $g_{bkc}$.
In \eqref{con:phig1}, if $g_{blc}> g_{bkc}$, $\phi_{bklc}=1$, otherwise, $\phi_{bklc}=0$.
Constraints \eqref{con:phi1} indicate that only one decoding order exists for each timeslot, e.g., either $k$ decoding $l$, or $l$ decoding $k$.

$\mathcal{P}_0$ is a Mixed-Integer Non-Convex Programming (MINCP) due to the binary variables, $\alpha_{bkc}$ and $\phi_{bklc}$, and the non-convexity of $R_{bkc}$.
Solving MINCP is in general challenging.
A typical way to address a max-min problem is to check whether it can be reformulated as a Monotonic Constrained Max-min Utility (MCMU) problem, where the objective functions and constraints are Competitive Utility Functions (CUFs) and Monotonic Constraints (MCs), respectively \cite{tan2015wireless}.
If yes, PF can be applied with fast convergence. 
The general MCMU is expressed as:
\begin{subequations}
\begin{align}
    \mathcal{P}_{PF}:&\max_{\mathbf{Q}}\,\,\,\,\min_{j=1,\dots,J}\,\,\,\,  f_j(\mathbf{Q}) \label{OPTpf}\\
    \,\,\,\,\mbox{s.t.}\,\,\,\,
    &F_m(\mathbf{Q})\leq \bar{F}_m, m=1,\dots,M.\label{con:f}
\end{align}
\end{subequations}
In $\mathcal{P}_{PF}$, $\mathbf{Q}=[Q_1,\dots,Q_j,\dots,Q_J]$ is the vector collecting all the $Q$-variables.
$f_j(\mathbf{Q})$ represents the objective function.
$F_m(\mathbf{Q})$ and $\bar{F}_m$ are the constraint functions and upper-bound parameters, respectively.
The properties of CUF and MC are presented in \textbf{Definition 1} and \textbf{Definition 2}, respectively.

\textbf{Definition 1.} \textit{The objective function $f_j(\mathbf{Q})$ in $\mathcal{P}_{PF}$ is CUF if the following properties are satisfied:}

\begin{itemize}
\item
\textit{Positivity: $f_j(\mathbf{Q})>0$ if $\mathbf{Q}\succ\mathbf{0}$; $f_j(\mathbf{Q})=0$ if and only if $\mathbf{Q}=\mathbf{0}$.}
\item
\textit{Competitiveness: $f_j(\mathbf{Q})$ strictly monotonically increases in $Q_j$ but decreases in $Q_{j^{'}}$, where $j^{'}\neq j$.}
\item
\textit{Directional Monotonicity: For $\zeta>1$ and $\mathbf{Q}\succ\mathbf{0}$, $f_j(\zeta\mathbf{Q})>f_j(\mathbf{Q})$.}
\end{itemize}

\textbf{Definition 2.} \textit{The constraints, $F_m(\mathbf{Q})\leq\bar{F}_m$, $\forall m=1,\dots,M$, are MCs if the following properties are satisfied:}

\begin{itemize}
\item
\textit{Strict Monotonicity: $F_m(\mathbf{Q}_1)>F_m(\mathbf{Q}_2)$ if $\mathbf{Q}_1\succ\mathbf{Q}_2$, $\forall m$.}
\item
\textit{Validity: If $\mathbf{Q}\succ \mathbf{0}$, $\exists\zeta>0$ such that $F_m(\zeta\mathbf{Q})\geq\bar{F}_m$ for some $m$.}
\end{itemize}

MCMU and PF may not be directly applied to solve $\mathcal{P}_0$ due to the following reasons:
\begin{itemize}
\item The solutions for MCMU (e.g., \cite{tan2015wireless,huang2013joint,zheng2017max}) are derived for a specific scenario, e.g., one terminal per cell or per beam.
When the scenario of multiple users per beam, along with undetermined decoding orders and binary variables, is considered in this paper, the satisfiability of \textbf{Definition 1} and \textbf{Definition 2} no longer holds for original $\mathcal{P}_0$.
\item In $\mathcal{P}_0$, determining decoding orders is coupled with beam power allocation.
Optimizing beam power could result in changes of decoding orders. 
As a consequence, the function of $R_{bk}$ in $\mathcal{P}_0$ becomes undetermined (corresponding to the objective function in $\mathcal{P}_{PF}$), which is an obstacle in analyzing the applicability of MCMU and PF.
\item Precoding vectors are decided based on the terminal-timeslot assignment.
The coupling between precoding vectors and terminal-timeslot assignment could result in undetermined $\vert\mathbf{h}_{bk}\mathbf{w}_{bc}\vert^2$ in the objective function \eqref{OPT1} while optimizing $\alpha_{bkc}$.
\end{itemize}

To solve $\mathcal{P}_0$, the following issues should be tackled.
First, the applicability of MCMU and PF for different special cases of $\mathcal{P}_0$ should be analyzed.
Second, the challenges to deal with the combinatorial and non-convex components in $\mathcal{P}_0$ need to be addressed. 
Towards these ends, we first discuss the optimization of power allocation and decoding orders with the fixed terminal-timeslot assignment.
Then we focus on solving the whole joint optimization problem.

\section{Optimal Joint Optimization of Power Allocation and Decoding Orders}
With fixed $\alpha_{bkc}$ in $\mathcal{P}_0$, we formulate the remaining power and decoding-order optimization problem in $\mathcal{P}_1$. 
\begin{subequations}
\begin{align}
    \mathcal{P}_1:&\max_{p_{bkc}>0,{\phi}_{bklc}}\,\,\,\,\min_{b\in\mathcal{B},k\in\mathcal{K}_b}\,\,\,\,  \dfrac{R_{bk}}{D_{bk}} \label{OPT2}\\
    \,\,\,\,\mbox{s.t.}\,\,\,\,
    &\eqref{con:pt1},\eqref{con:pbm1},\eqref{con:pc1},\eqref{con:phig1},\eqref{con:phi1}.
\end{align}
\end{subequations}
Note that prior to optimization, we have pre-processed $p_{bkc}$ according to the fixed variables $\alpha_{bkc}$.
That is, only positive $p$-variables, i.e., $p_{bkc}>0$ (resulted by $\alpha_{bkc}=1$), retain in $\mathcal{P}_1$ and to be optimized.
$\mathcal{P}_1$ is complicated due to the coupled power and decoding-order optimization.
From $\mathcal{P}_1$, we can observe that if the decoding orders can be determined by temporarily fixing the beam power, the remaining power allocation problem resembles $P_{PF}$.
This enables us to take advantages of the PF method in fast convergence and optimality guarantee.
In this section, we first discuss  the strategy of fixing the terminal-timeslot assignment.
Next, we discuss the solution of $\mathcal{P}_1$, and the applicability of MCMU and PF.

\subsection{Terminal-Timeslot Scheduling}
Terminal-timeslot scheduling or terminal grouping is significant for NOMA and precoding.
In the literature, the grouping strategies are either optimal or suboptimal.
The former is to find the optimal terminal groups but with prohibitively computational complexity, e.g., an optimal scheme for joint precoding and terminal-subcarrier assignment in \cite{sun2017optimal}.
For the latter, some heuristic approaches are developed for terrestrial-NOMA systems but might not be directly applied to satellite NOMA. 
For example, the strategy of grouping terminals with highly correlated channels and large channel gain difference is widely used in terrestrial-NOMA systems \cite{liu2017joint, chinnadurai2017novel,ali2016non}.
However, in satellite systems, neighboring terminals may have highly correlated channels but small channel gain difference \cite{perez2019non}, whereas terminals far away from each other may have non-correlated channels.

Considering the trade-off between interference reduction and computational complexity, we apply MaxCC strategy to select terminals with the largest correlation \cite{christopoulos2015multicast}.
The reasoning behind this strategy is that the precoder should be able to mitigate inter-beam interference more effectively whenever the terminals grouped with the same beam have highly correlated channel vectors.
The procedure is summarized in the following.
In a timeslot, we select one terminal, say $k^{'}$, randomly from $\mathcal{U}_b$.
Then we calculate its correlation factors (or cosine similarity metric) with all the other terminals, i.e., $\theta=\frac{\vert\mathbf{h}_{bk^{'}}^H\mathbf{h}_{bj}\vert}{\Vert\mathbf{h}_{bk^{'}}\Vert\Vert\mathbf{h}_{bj}\Vert}$ \cite{caus2016noma}, where $j\in\mathcal{U}_b\setminus\lbrace k^{'}\rbrace$.
The terminal with the largest $\theta$ is scheduled with $k^{'}$ to the same timeslot.
The selected terminals are deleted from $\mathcal{U}_b$ and added to $\mathcal{K}_b$.
The above procedure is performed for each timeslot one by one until all the timeslots are processed or $\mathcal{U}_b$ becomes empty.

\subsection{Terminal Power Optimization with Fixed Beam Power}
We define $\mathbf{P}=[P_1,\dots,P_b,\dots,P_B]$ as the vector collecting all the beam power.
With fixed $\alpha_{bkc}$ and temporarily fixed $\mathbf{P}$, the terminal power allocation is independent among beams.
Thus $\mathcal{P}_1$ can be decomposed to $B$ subproblems.
The $b$-th subproblem, $\mathcal{P}_1(b)$, corresponds to the terminal power optimization in beam $b$.
Let $\bar{\mathbf{P}}_b$ collect all the beam power except the $b$-th beam's power, i.e., $\bar{\mathbf{P}}_b=[P_1,\dots,P_{b-1},P_{b+1},\dots,P_B]$.
In \eqref{equ:g}, $g_{bkc}$ can be considered as a function of $\bar{\mathbf{P}}_b$, which is defined as,
\begin{equation}
g_{bkc}=\hat{f}_{bkc}(\bar{\mathbf{P}}_b).
\end{equation}
The decoding order variables $\phi_{bklc}$ are determined when $\mathbf{P}$ is fixed.
Thus, constraints \eqref{con:phig1} and \eqref{con:phi1} do not apply in $\mathcal{P}_1(b)$.
\begin{subequations}
\begin{align}
    \mathcal{P}_1(b):&\max_{p_{bkc}}\,\,\,\,\min_{k\in\mathcal{K}_b}\,\,\,\,  \dfrac{R_{bk}}{D_{bk}} \label{OPT2b}\\
    \,\,\,\,\mbox{s.t.}\,\,\,\,
    &\rho_{bc}\sum_{k\in\mathcal{K}_b}p_{bkc}= P_{b},\forall c\in\mathcal{C},\label{con:pc2b}
\end{align}
\end{subequations}
where \eqref{con:pc1} is equivalently converted to \eqref{con:pc2b} and denotes that the sum of terminals' power in each beam across timeslots is equal to the beam power.
By introducing an auxiliary variable $t_b$, $\mathcal{P}_1(b)$ can be equivalently transformed to a maximization problem:
\begin{subequations}
\begin{align}
    \mathcal{P}_1(b):&\max_{p_{bkc},t_b}\,\,\,\,t_b \label{OPT2be}\\
    \,\,\,\,\mbox{s.t.}\,\,\,\,
    &\eqref{con:pc2b},\\
    &t_bD_{bk}-R_{bk}\leq 0, \forall k\in\mathcal{K}_b.\label{con:rd1}
\end{align}
\end{subequations}
To better reveal the convexity of $\mathcal{P}_1(b)$, we express $p_{bkc}$ by a function of $R_{bkc}$ based on \eqref{equ:r} \cite{you2018resource}.
Then the power variables $p_{b1c},\dots,p_{bK_bc}$ read:
\begin{align}
p_{b1c}&=\dfrac{e^{\frac{R_{b1c}}{B_W}}-1}{g_{b1c}},\notag\\
p_{b2c}&=\dfrac{e^{\frac{R_{b2c}}{B_W}}-1}{g_{b2c}}(g_{b2c}p_{b1c}+1),\notag\\
&\vdots\notag\\
p_{bK_bc}&=\dfrac{e^{\frac{R_{bK_bc}}{B_W}}-1}{g_{bK_bc}}\left (g_{bK_bc}\sum_{j=1}^{K_b-1}p_{bjc}+1\right ).
\end{align}
The constraints in \eqref{con:pc2b} can be equivalently written as:
\begin{equation}
\sum_{k=1}^{K_b}\left(\dfrac{1}{g_{bkc}}-\dfrac{1}{g_{b(k-1)c}}\right) e^{\sum\limits_{j\geq k} \frac{R_{bjc}}{B_W}}-\dfrac{1}{g_{bK_bc}}=\dfrac{P_{b}}{\rho_{bc}},\forall c\in\mathcal{C},\label{con:pt3bc1}
\end{equation}
where $\frac{1}{g_{b0c}}=0$.
Then $\mathcal{P}_1(b)$ is equivalently converted to $\mathcal{P}_2(b)$ by treating ${R}_{bkc}$ as variables:
\begin{subequations}
\begin{align}
    \mathcal{P}_2(b):&\max_{{R}_{bkc},t_b}\,\,\,\,t_b \label{OPT3bc1}\\
    \,\,\,\,\mbox{s.t.}\,\,\,\,
    &\eqref{con:pt3bc1},\eqref{con:rd1}. 
\end{align}
\end{subequations}
Note that constraints \eqref{con:pt3bc1} are not affine. We further relax the equality constraints in \eqref{con:pt3bc1} to inequality in \eqref{con:pt2}, leading to a convex exponential-cone format,
\begin{equation}
\sum_{k=1}^{K_b}\left(\dfrac{1}{g_{bkc}}-\dfrac{1}{g_{b(k-1)c}}\right) e^{\sum\limits_{j\geq k} \frac{R_{bjc}}{B_W}}-\dfrac{1}{g_{bK_bc}}\leq\dfrac{P_{b}}{\rho_{bc}},\forall c\in\mathcal{C}.\label{con:pt2}
\end{equation}
We then conclude the equivalence between \eqref{con:pt3bc1} and \eqref{con:pt2} at the optimum, thus concluding the convexity of $\mathcal{P}_2(b)$ and $\mathcal{P}_1(b)$.

\textbf{Proposition 1.} \textit{The optimum of $\mathcal{P}_2(b)$, i.e., $t_b^*$, which is located on timeslot $c^*$, can be obtained by the following equation:}
\begin{equation}
\sum_{k=1}^{K_b}\left( \dfrac{1}{g_{bkc^*}}-\dfrac{1}{g_{b(k-1)c^*}}\right) e^{\sum\limits_{j\geq k} \frac{t_{b}^*D_{bj}}{B_W}}-\dfrac{1}{g_{bK_bc^*}} = \frac{P_{b}}{\rho_{bc^*}}.
\label{equ:tpc0}
\end{equation}

\begin{proof}
We can obtain the optimum of the relaxed problem based on Karush-Kuhn-Tucker (KKT) conditions.
The corresponding Lagrangian dual function is:
\begin{align}
&\mathcal{L}({R}_{bkc},t_b;{\lambda}_c,{\mu}_k)=-t_b\notag\\
&+\sum_{c\in\mathcal{C}}\lambda_c \left (\sum_{k=1}^{K_b}\left (\frac{1}{g_{bkc}}-\frac{1}{g_{b(k-1)c}}\right ) e^{\sum\limits_{j\geq k}\frac{R_{bjc}}{B_W}}-\frac{1}{g_{bK_bc}}-\frac{P_{b}}{\rho_{bc}} \right )\notag\\
&+\sum_{k=1}^{K_b}\mu_{k}(tD_{bk}-R_{bk}),
\end{align}
\noindent where ${\lambda}_c\geq 0$ and ${\mu}_k\geq 0$ are Lagrangian multipliers for constraints \eqref{con:pt3bc1} and \eqref{con:rd1}, respectively.
The KKT conditions can be derived as
\begin{subequations}
\begin{align}
&\frac{\partial \mathcal{L}}{\partial R_{bkc}}=\lambda_c \sum_{n=1}^{k}\left(\dfrac{1}{g_{bnc}}-\dfrac{1}{g_{b(n-1)c}}\right) e^{ \sum\limits_{j\geq n}\frac{R_{bjc}}{B_W}}-\mu_{k}=0,\notag \\
&\,\,\,\,\,\,\,\,\,\,\,\,\,\,\,\,\,\,\,\,\,\,\,\,\,\,\,\,\,\,\,\,\,\,\,\,\,\,\,\,\,\,\,\,\,\,\,\,\,\,\,\,\,\,\,\,\,\,\,\,\,\,\,\,\,\,\,\,\,\,\,\,\,\,\,\,\,\,\,\,\,\,\,\,\,\,\,\,\,\,\,\,\,\forall c\in\mathcal{C},k\in\mathcal{K}_b,\label{con:kkt12}\\
&\frac{\partial \mathcal{L}}{\partial t}=-1+\sum_{k=1}^{K_b}\mu_{k}D_{bk}=0,\label{con:kkt22}\\
&\lambda_c\left(\sum_{k=1}^{K_b}\left( \dfrac{1}{g_{bkc}}-\dfrac{1}{g_{b(k-1)c}}\right) e^{\sum\limits_{j\geq k} \frac{R_{bjc}}{B_W}}-\dfrac{1}{g_{bK_bc}} - \dfrac{P_{b}}{\rho_{bc}}\right)=0,\notag \\
&\,\,\,\,\,\,\,\,\,\,\,\,\,\,\,\,\,\,\,\,\,\,\,\,\,\,\,\,\,\,\,\,\,\,\,\,\,\,\,\,\,\,\,\,\,\,\,\,\,\,\,\,\,\,\,\,\,\,\,\,\,\,\,\,\,\,\,\,\,\,\,\,\,\,\,\,\,\,\,\,\,\,\,\,\,\,\,\,\,\,\,\,\,\,\,\,\,\,\,\,\,\,\,\,\,\,\,\,\,\,\,\,\,\,\forall c\in\mathcal{C},\label{con:kkt23}\\
&\mu_{k} (t_bD_{bk}-R_{bk})=0,\forall k\in\mathcal{K}_b.\label{con:kkt24}
\end{align}
\end{subequations}
At the optimum of $\mathcal{P}_1(b)$, at least one constraint in \eqref{con:rd1}, say the $k^*$-th constraint/terminal, will be active, i.e., the equality holds, whereas the others keep inequalities \cite{boyd2004convex}.
The optimal value $t_b^*$ is then achieved at the equality $t_b^*D_{bk^*}-R_{bk^*}=0$ \cite{zheng2017max, boyd2004convex}.
In \eqref{con:kkt24}, for the inequality terms $t_b^*D_{bk}-R_{bk}<0$, the corresponding $\mu_k$ must be zero, while for the equality term $t_b^*D_{bk^*}-R_{bk^*}=0$, $\mu_{k^*}>0$ instead of zero since \eqref{con:kkt22} cannot hold for all-zero $\mu_k$.
Hence, the optimal $t_b^*$ is associated with positive $\mu_k^*$.
The positive $\mu_k^*$ in \eqref{con:kkt12} results in positive $\lambda_c$ which leads to $\sum_{k=1}^{K_b}( \frac{1}{g_{bkc}}-\frac{1}{g_{b(k-1)c}}) e^{\sum_{j\geq k} \frac{R_{bjc}}{B_W}}-\frac{1}{g_{bK_bc}} - \frac{P_{b}}{\rho_{bc}}=0$ in \eqref{con:kkt23}.
Thus the conclusion.
\end{proof}

\textbf{Proposition 1} establishes the equivalence between \eqref{con:pt3bc1} and  \eqref{con:pt2} at the optimum.
The convexity of $\mathcal{P}_1(b)$ and $\mathcal{P}_2(b)$ is concluded.
We define a function $t_b^*=f_b(\mathbf{P})$ in an inexplicit way in \eqref{equ:tpc0} by moving $t_b^*$ to the left side of the equality and the remaining to the right, where $f_{b}(\mathbf{P})$ denotes the function of the optimal OCTR of the $b$-th beam when beam power is $\mathbf{P}$.

\subsection{Beam Power Optimization}
Given $\mathbf{P}$, the optimal power allocation among terminals can be obtained from KKT conditions.
Next, we optimize the beam power allocation.
The problem is formulated in $\mathcal{P}_3$, 
\begin{subequations}
\begin{align}
    \mathcal{P}_3:&\max_{\mathbf{P}}\,\,\,\,\min_{b\in\mathcal{B}}\,\,\,\, f_b(\mathbf{P})  \label{OPT5b}\\
    \,\,\,\,\mbox{s.t.}\,\,\,\,
    &\sum_{b\in\mathcal{B}}P_{b}\leq P_{tot},\label{con:pt5b}\\
    &P_{b}\leq P_{b,\max},\forall b\in\mathcal{B},\label{con:pbm5b}
\end{align}
\end{subequations}
where the objective $f_b(\mathbf{P})$ is the function of the optimal OCTR of the $b$-th beam with $\mathbf{P}$ and can be equivalently converted from \eqref{equ:tpc0}.
The expression of $f_b(\mathbf{P})$ depends on $\mathbf{P}$ and the decoding order.
Next, we prove $\mathcal{P}_3$ is an MCMU.
Constraints \eqref{con:pt5b} and \eqref{con:pbm5b} are linear, which satisfy the MC conditions.
The CUF conditions of $f_b(\mathbf{P})$ are analyzed in \textbf{Lemma 1} and \textbf{Lemma 2}.

\textbf{Lemma 1.} \textit{The objective function $f_b(\mathbf{P})$ in $\mathcal{P}_3$ is a CUF for any decoding orders.}
\begin{proof}
Given any $\mathbf{P}$ and the corresponding decoding order, according to \textbf{Definition 1}, we check the three conditions for $f_{b}(\mathbf{P})$, $\forall b\in\mathcal{B}$.

\textit{Positivity}: 
Rewrite \eqref{equ:tpc0} equivalently as:
\begin{align}
\sum_{k=1}^{K_b-1}\dfrac{1}{g_{bkc^*}} e^{\sum\limits_{j>k} \frac{t_{b}^*D_{bj}}{B_W}}(e^{\frac{t^*_bD_{bk}}{B_W}}-1)\notag\\
+\dfrac{1}{g_{bK_bc^*}}(e^{\frac{t^*_bD_{bK_b}}{B_W}}-1)= \frac{P_{b}}{\rho_{bc^*}}.
\label{equ:tpc1}
\end{align}
The right-hand side is positive, then the term $e^{\frac{t^*_bD_{bk}}{B_W}}-1$ in the left-hand side has to keep positive. Hence $t_b^*$ is positive.

\textit{Competitiveness}: 
By deriving the partial derivatives of $f_{b}(\mathbf{P})$, i.e., $\frac{\partial f_{b}}{\partial P_b}$ and $\frac{\partial f_{b}}{\partial P_{b^{'}}}$ in \eqref{equ:dfp1} and \eqref{equ:dfp2}, respectively,  we observe $\frac{\partial f_{b}}{\partial P_b}>0$ and $\frac{\partial f_{b}}{\partial P_{b^{'}}}<0$, which means that $f_{b}(\mathbf{P})$ monotonically increases with beam $b$'s power $P_b$ but decreases with any other beam's power $P_{b^{'}}$.
\begin{figure*}[!t]
\normalsize
\begin{equation}
\label{equ:dfp1}
\frac{\partial f_{b}}{\partial P_b}=\frac{1}{\sum_{k=1}^{K_b}\left(\frac{1}{g_{bkc^*}}-\frac{1}{g_{b(k-1)c^*}}\right)e^{\sum_{j\geq k}\frac{t_{b}^*D_{bj}}{B_W}}\sum_{j\geq k}\frac{D_{bj}}{B_W}},
\end{equation}
\begin{equation}
\label{equ:dfp2}
\frac{\partial f_{b}}{\partial P_{b^{'}}}=-\frac{\sum_{k=1}^{K_b-1}\frac{\vert\mathbf{h}_{bk}^H\mathbf{w}_{b^{'}c^*}\vert^2}{\vert\mathbf{h}_{bk}^H\mathbf{w}_{bc^*}\vert^2\rho_{bc^*}}e^{\sum_{j> k}\frac{t_{b}^*D_{bj}}{B_W}}\left(e^{\frac{t_{b}^*D_{bk}}{B_W}}-1\right)+\frac{\vert\mathbf{h}_{bK_b}^H\mathbf{w}_{b^{'}c^*}\vert^2}{\vert\mathbf{h}_{bK_b}^H\mathbf{w}_{bc^*}\vert^2\rho_{bc^*}}\left(e^{\frac{t_{b}^*D_{bK_b}}{B_W}}-1\right)}{\sum_{k=1}^{K_b}\left(\frac{1}{g_{bkc^*}}-\frac{1}{g_{b(k-1)c^*}}\right)e^{\sum_{j\geq k}\frac{t_{b}^*D_{bj}}{B_W}}\sum_{j\geq k}\frac{D_{bj}}{B_W}}.
\end{equation}
\hrulefill
\end{figure*}

\textit{Directional Monotonicity:}
Let $\zeta>1$.
We assume $f_b(\zeta\mathbf{P})=\tau_1$ and $f_b(\mathbf{P})=\tau_2$.
From equation \eqref{equ:tpc0}, $\tau_1$ can be derived by the following equation:
\begin{align}
\sum_{k=1}^{K_b-1}\dfrac{1}{\hat{f}_{bkc^*}(\zeta\bar{\mathbf{P}}_b)}e^{\sum\limits_{j> k} \frac{\tau_1 D_{bj}}{B_W}}(e^{\frac{\tau_1 D_{bk}}{B_W}}-1)\notag\\+\dfrac{1}{\hat{f}_{bK_bc^*}(\zeta\bar{\mathbf{P}}_b)}(e^{\frac{\tau_1 D_{bK_b}}{B_W}}-1) - \frac{\zeta P_{b}}{\rho_{bc^*}}=0.
\label{equ:ltp1}
\end{align}
By substituting $f_b(\mathbf{P})=\tau_2$ into \eqref{equ:tpc0}, both sides of the equation multiply $\zeta$, i.e., 
\begin{align}
\sum_{k=1}^{K_b-1}\dfrac{\zeta}{\hat{f}_{bkc^*}(\bar{\mathbf{P}}_b)}e^{\sum\limits_{j> k} \frac{\tau_2 D_{bj}}{B_W}}(e^{\frac{\tau_2 D_{bk}}{B_W}}-1)\notag\\+\dfrac{\zeta}{\hat{f}_{bK_bc^*}(\bar{\mathbf{P}}_b)}(e^{\frac{\tau_2 D_{bK_b}}{B_W}}-1) - \frac{\zeta P_{b}}{\rho_{bc^*}}=0.
\label{equ:ltp2}
\end{align}
Based on the equation in \eqref{equ:g}, we can derive $\frac{1}{\hat{f}_{bkc^*}(\zeta\bar{\mathbf{P}})}<\frac{\zeta}{\hat{f}_{bkc^*}(\bar{\mathbf{P}})}$ by:
\begin{align}
&\frac{1}{\hat{f}_{bkc^*}(\zeta\bar{\mathbf{P}})}=\frac{\sum_{b^{'}\neq b}\vert\mathbf{h}_{bk}^H\mathbf{w}_{b^{'}c^*}\vert^2\zeta \frac{P_{b^{'}}}{\rho_{b^{'}c^*}}+\sigma^2}{\vert\mathbf{h}_{bk}^H\mathbf{w}_{bc^*}\vert^2}\notag\\
&\,\,\,\,\,\,\,\,\,\,\,\,\,\,\,<\zeta\frac{\sum_{b^{'}\neq b}\vert\mathbf{h}_{bk}^H\mathbf{w}_{b^{'}c^*}\vert^2\frac{P_{b^{'}}}{\rho_{b^{'}c^*}}+\sigma^2}{\vert\mathbf{h}_{bk}^H\mathbf{w}_{bc^*}\vert^2}=\frac{\zeta}{\hat{f}_{bkc^*}(\bar{\mathbf{P}}_b)}.
\label{equ:gzp}
\end{align}
Based on \eqref{equ:gzp}, the terms $\frac{1}{\hat{f}_{bkc^*}(\zeta\bar{\mathbf{P}})}$ and $\frac{1}{\hat{f}_{bK_bc^*}(\zeta\bar{\mathbf{P}})}$ in \eqref{equ:ltp1} are smaller than $\frac{\zeta}{\hat{f}_{bkc^*}(\bar{\mathbf{P}}_b)}$ and $\frac{\zeta}{\hat{f}_{bK_bc^*}(\bar{\mathbf{P}}_b)}$ in \eqref{equ:ltp2}, respectively.
Hence, the equalities in \eqref{equ:ltp1} and \eqref{equ:ltp2} cannot hold under both cases $\tau_1=\tau_2$ and $\tau_1<\tau_2$.
Thus $\tau_1>\tau_2$ and $f_b(\zeta\mathbf{P})>f_b(\mathbf{P})$.
\end{proof}

Based on \textbf{Lemma 1}, we can develop PF-based algorithms to converge if the decoding order remains under the power adjustment.
However, the expressions of $f_b(\mathbf{P})$ typically change since the adjustment of $\mathbf{P}$ can result in new decoding orders. 
As a consequence, it is not straightforward to observe the satisfiability of CUF and the convergence when $f_b(\mathbf{P})$ varies.
Next, we conclude that the objective function in $\mathcal{P}_3$ is a CUF even if the decoding order changes.

\textbf{Lemma 2.} \textit{$f_b(\mathbf{P})$ in $\mathcal{P}_3$ remains a CUF even if the decoding order changes.}
\begin{proof}
\textit{Positivity}: 
The positivity of $f_b(\mathbf{P})$ holds whether the decoding order changes or not according to \eqref{equ:tpc1}.

\textit{Competitiveness}: 
The decoding order in beam $b$ depends on $\bar{\mathbf{P}}_b$.
Given any two terminals $k$ and $k^{'}$ in beam $b$, suppose that in beam $b^{'}$, there exist $P_{b^{'}}$ and $\delta$ such that $P_{b^{'}}$ leads to $g_{bkc}=g_{bk^{'}c}$; setting $P_{b^{'}}-\delta$ results in terminal $k$ decoding $k^{'}$ ($g_{bkc}>g_{bk^{'}c}$); and $P_{b^{'}}+\delta$ changes the decoding order to $k^{'}$ decoding $k$ ($g_{bkc}<g_{bk^{'}c}$).
$f_{b}({\mathbf{P}})$ is competitive when the decoding order stays unchanged.
When $g_{bkc}=g_{bk^{'}c}$, $f_{b}({\mathbf{P}})$ remains the same under both decoding orders.
Thus $f_{b}(\mathbf{P})$ is continuous, indicating that $f_{b}({\mathbf{P}})$ monotonically decreases in $P_{b^{'}}$ even if the decoding order changes.
The competitiveness is concluded.

\textit{Directional monotonicity}: 
Assume that the decoding order changes from $k$ decoding $k^{'}$ to $k^{'}$ decoding $k$ as the beam power increases from $\mathbf{P}$ to $\zeta\mathbf{P}$, where $\zeta>1$.
There exists $\zeta_0$, where $1<\zeta_0<\zeta$, such that $\zeta_0\mathbf{P}$ corresponds to $g_{bkc}=g_{bk^{'}c}$.
As proven in \textbf{Lemma 1}, $f_b(\mathbf{P})<f_b(\zeta_0\mathbf{P})$ and $f_b(\zeta_0\mathbf{P})<f_b(\zeta\mathbf{P})$.
Thus $f_b(\mathbf{P})<f_b(\zeta\mathbf{P})$.
\end{proof}

Based on \textbf{Lemma 1} and \textbf{Lemma 2}, the objective function $f_b(\mathbf{P})$ in $\mathcal{P}_3$ is a CUF.
Constraints \eqref{con:pt5b} and \eqref{con:pbm5b} are linear and thus satisfy the MC conditions, which concludes $\mathcal{P}_3$ is an MCMU.

\subsection{A Fast-Convergence Approach Based on PF for Joint Power and Decoding-Order Optimization}

\begin{algorithm}[t]
\caption{JOPD}
\label{alg:2}
\begin{algorithmic}[1]
\REQUIRE ~~\\
Initial beam power: $\mathbf{P}^{(0)}$,\\ Iteration index: $n=0$.
\REPEAT 
\STATE Update and sort $g_{bkc}$ with $\mathbf{P}^{(n)}$. 
\STATE Determine decoding order ${\phi}_{bklc}$ based on the descending order of $g_{bkc}$.
\STATE Calculate $t_b^{*(n)}=f_b(\mathbf{P}^{(n)})$, $\forall b\in\mathcal{B}$, by \eqref{equ:tpc0}.
\STATE Update $\mathbf{P}$ by $P_{b}=\frac{P_{b}^{(n)}}{t_b^{*(n)}}$, $\forall b\in\mathcal{B}$.
\STATE Calculate $\epsilon=\max\left\{ \frac{P_{b}}{P_{b,\max}}, \forall b\in\mathcal{B}; \sum_{b\in\mathcal{B}}\frac{P_{b}}{P_{tot}}\right\}$.
\STATE Update $\mathbf{P}^{(n+1)}=\frac{\mathbf{P}}{\epsilon}$, $n=n+1$.
\UNTIL{convergence}
\STATE Calculate $p_{bkc}$ based on $\mathbf{P}^{(n)}$.
\ENSURE ~~\\
$t^*_b$,  $p_{bkc}$.
\end{algorithmic}
\end{algorithm}

$\mathcal{P}_3$ is an MCMU where the objective function is CUF and the constraints are MCs.
We propose an iterative algorithm based on PF, i.e., JOPD, in Alg. \ref{alg:2} to solve $\mathcal{P}_3$.
Let $\mathbf{P}^{(n)}$, $P_b^{(n)}$ and $t_{b}^{*(n)}$ represent the values of $\mathbf{P}$, $P_b$ and $t^*_b$ at the $n$-th iteration, respectively. 
For each iteration, decoding orders are updated according to the descending order of $g_{bkc}$ in line 2 and line 3.
Then the optimal OCTR of each beam is calculated in line 4.
Beam power is adjusted inversely proportional to the value of $t^*_b$ in line 5 \cite{tan2015wireless}, which suggests that power for the beams with larger $t_b^*$ will be reduced in the next iteration, and more power is allocated to the beams with worse OCTR.
In line 6 and line 7, we introduce a factor $\epsilon$ to confine beam power in the domain of \eqref{con:pt5b} and \eqref{con:pbm5b}.
The convergence and optimality of JOPD are concluded in \textbf{Theorem 1}.

\textbf{Theorem 1.} 
\textit{With any initial vector $\mathbf{P}$, JOPD converges geometrically fast to the optimum of $\mathcal{P}_3$.}

\begin{proof}
At the optimum, $f_b(\mathbf{P}^*)=t^*$, $\forall b\in\mathcal{B}$, where $\mathbf{P}^*=[P_1^*,\dots,P_b^*,\dots,P_B^*]$ and $t^*$ are the optimal beam power and the optimal OCTR value, respectively.
Define function $\eta_b(\mathbf{P})=\frac{P_b}{f_{b}(\mathbf{P})},\forall b\in\mathcal{B}$.
At the convergence, $\frac{P_b^*}{t^*}=\frac{P_b^*}{f_b(\mathbf{P}^*)}$, $\forall b\in\mathcal{B}$.

The algorithm converges geometrically fast to $t^*$ with any initial $\mathbf{P}$ if $\eta_b(\mathbf{P})$ satisfies the following conditions \cite{zheng2016wireless}:
\begin{itemize}
\item There exist $\underline{\tau}$ and $\overline{\tau}$, where $0<\underline{\tau}\leq\overline{\tau}$, such that $\underline{\tau}\leq \eta_b(\mathbf{P})\leq\overline{\tau}$, $\forall b\in\mathcal{B}$. 
\item For any beam power $\mathbf{P}_1\succ \mathbf{0}$ and $\mathbf{P}_2\succ \mathbf{0}$, and $0<\zeta\leq 1$, if $\zeta\mathbf{P}_1\preceq \mathbf{P}_2$, then $\zeta \eta_b(\mathbf{P}_1)\leq \eta_b(\mathbf{P}_2)$, $\forall b\in\mathcal{B}$. For $0<\zeta<1$, if $\zeta\mathbf{P}_1\prec\mathbf{P}_2$, then $\zeta \eta_b(\mathbf{P}_1)< \eta_b(\mathbf{P}_2)$, $\forall b\in\mathcal{B}$.
\end{itemize}

For the first condition, $\eta_{b}(\mathbf{P})$ stays between $\underline{\tau}$ and $\overline{\tau}$, which means the function could not be zero or infinite with any $\mathbf{P}$.
Due to the positivity of $f_b(\mathbf{P})$, $\eta_b(\mathbf{P})=\frac{P_b}{f_b(\mathbf{P})}>0$, i.e., $\eta_b(\mathbf{P})\geq \underline{\tau}>0$.
Since $\mathbf{P}$ is bounded by $P_{b,\max}$, $\eta_b(\mathbf{P})$ is finite.
Thus the function is upper bounded, i.e., $\eta_b(\mathbf{P})\leq\overline{\tau}$.

For the second condition, we prove $\zeta \eta_b(\mathbf{P}_1)\leq \eta_b(\mathbf{P}_2)$ via showing the inequality below.
\begin{equation}
\zeta \eta_b(\mathbf{P}_1)\leq\eta_{b}(\zeta\mathbf{P}_1)\leq\eta_b(\mathbf{P}_2).
\label{equ:rela}
\end{equation}
The first inequality $\zeta \eta_b(\mathbf{P}_1)\leq\eta_{b}(\zeta\mathbf{P}_1)$ reads,
\begin{equation}
\frac{\zeta P_b}{f_b(\mathbf{P}_1)}\leq\frac{\zeta P_b}{f_{b}(\zeta\mathbf{P}_1)}.
\label{equ:rela1}
\end{equation}
Let $\zeta\mathbf{P}_1=\mathbf{P}$, then $\mathbf{P}_1=\frac{1}{\zeta}\mathbf{P}$, where $\frac{1}{\zeta}\geq 1$.
According to \textbf{Lemma 1} and \textbf{Lemma 2}, $f_b(\mathbf{P})$ satisfies directional monotonicity, thus $f_b(\frac{1}{\zeta}\mathbf{P})\geq f_b(\mathbf{P})$ and $\zeta \eta_b(\mathbf{P}_1)\leq\eta_{b}(\zeta\mathbf{P}_1)$ holds.
For the second inequality in \eqref{equ:rela}, $\eta_b(\zeta\mathbf{P}_1)\leq \eta_b(\mathbf{P}_2)$.
Based on $\frac{\partial f_{b}}{\partial P_{b}}>0$ and $\frac{\partial f_{b}}{\partial P_{b^{'}}}<0$ in \eqref{equ:dfp1} and \eqref{equ:dfp2}, we can derive the partial derivatives of $\eta_b(\mathbf{P})$ as:
\begin{equation}
\frac{\partial \eta_{b}}{\partial P_b}=\frac{f_b(\mathbf{P})-P_b\frac{\partial f_{b}}{\partial P_b}}{f^2_b(\mathbf{P})},
\end{equation}
\begin{equation}
\frac{\partial \eta_{b}}{\partial P_{b^{'}}}=-\frac{P_b\frac{\partial f_{b}}{\partial P_{b^{'}}}}{f^2_b(\mathbf{P})},
\end{equation}
where $\frac{\partial \eta_{b}}{\partial P_{b^{'}}}$ is positive.
We derive $\frac{\partial^2 f_{b}}{\partial P_b^2}<0$ based on \eqref{equ:dfp1}, which indicates the concavity of $f_b(\mathbf{P})$ on $P_b$ \cite{boyd2004convex}.
Let $\mathbf{P}_0=[P_1,\dots,0,\dots,P_B]$.
According to the first-order condition of concavity \cite{boyd2004convex} and $f_b(\mathbf{P}_0)=0$, $f_b(\mathbf{P})-f_b(\mathbf{P}_0)>(P_b-0)\frac{\partial f_{b}}{\partial P_b}$, and thus $\frac{\partial \eta_{b}}{\partial P_b}=\frac{f_b(\mathbf{P})-P_b\frac{\partial f_{b}}{\partial P_b}}{f^2_b(\mathbf{P})}>0$.
The monotonicity of $\eta_b(\mathbf{P})$ is concluded, i.e., $\eta_b(\mathbf{P})$ is an increasing function of $\mathbf{P}$.
Hence $ \eta_b(\zeta\mathbf{P}_1)\leq \eta_b(\mathbf{P}_2)$ holds in \eqref{equ:rela}, and thus $\zeta \eta_b(\mathbf{P}_1)\leq \eta_b(\mathbf{P}_2)$.
The result that $\zeta \eta_b(\mathbf{P}_1)< \eta_b(\mathbf{P}_2)$ if $\zeta_0\mathbf{P}_1\prec\mathbf{P}_2$ follows analogously.
\end{proof}

Next, in \textbf{Corollary 1}, we conclude that although the optimal beam power, coupling with decoding orders, in $\mathcal{P}_1$ is challenging to be directly obtained, the optimum of $\mathcal{P}_1$, in fact, can be achieved by solving a simple problem, i.e., $\mathcal{P}_3$.

\textbf{Corollary 1.}
\textit{The optimum of $\mathcal{P}_1$ is equal to that of $\mathcal{P}_3$.}

The reasons can be explained as follows.
$\mathcal{P}_1$ and $\mathcal{P}_3$ solves de facto the same problem, i.e., with the fixed $\alpha$-variables then obtain the max-min OCTR along with the optimal beam and terminal power allocation since in $\mathcal{P}_3$, when $\mathbf{P}$ is known, $p_{bkc}$ is also known.
\textbf{Theorem 1} indicates that, under the same $\alpha_{bkc}$, no better beam power allocation than $\mathbf{P}^*$ can be found. 
Thus $\mathbf{P}^*$ is optimal for $\mathcal{P}_1$ and $\mathcal{P}_3$.
Given $\mathbf{P}^*$ to $\mathcal{P}_1$, the resulting max-min OCTR and terminal power allocation are therefore optimal, and thus the conclusion.

The difference between $\mathcal{P}_1$ and $\mathcal{P}_3$ is that, in $\mathcal{P}_1$, one has to deal with the issue of unconfirmed convergence and undetermined optimal $R_{bk}$ expressions due to the decoding-order variations and the undetermined optimal decoding order.
In $\mathcal{P}_3$, we circumvent these difficulties by using the established analytical results in this section.
By solving $\mathcal{P}_3$ via Alg. 1, we update beam power associated with decoding order successively, instead of obtaining the optimum directly.
Guaranteed by \textbf{Lemma 1}, \textbf{Lemma 2}, and \textbf{Theorem 1}, this simple power-adjustment approach eventually leads to the optimal beam power and optimal decoding order for the given $\alpha$-variables.

\section{A Heuristic Algorithm for Joint Power, Decoding-Order, and Terminal-Timeslot Optimization}
JOPD is limited by the one-off terminal-timeslot assignment.
Based on the framework of JOPD and taking its fast-convergence advantages, we design a heuristic approach, JOPDT, to iteratively update timeslot-terminal assignment and improve the overall performance.
The procedure of the heuristic approach is presented in Alg. 2.

Line 3 to line 10 present the process of implementing the JOPD framework.
In line 2 and line 5, precoding vectors and decoding orders are updated based on the terminal-timeslot assignment and beam power allocation, respectively.
In line 6, a joint power-allocation, decoding-order, and terminal-timeslot optimization problem is solved.
The problem is constructed as follows.
Analogous to JOPD, by fixing $\mathbf{P}$, $\mathcal{P}_0$ is decomposed into $B$ subproblems, each of which represents the optimization of terminals' power allocation and terminal-timeslot assignment in the beam.
The $b$-th subproblem is expressed as,
\begin{subequations}
\begin{align}
    \mathcal{P}_4(b):&\max_{p_{bkc},\alpha_{bkc}}\,\,\,\,\min_{k\in\mathcal{K}_b}\,\,\,\,  \dfrac{R_{bk}}{D_{bk}} \label{OPT6b}\\
    \,\,\,\,\mbox{s.t.}\,\,\,\,
    &\rho_{bc}\sum_{k\in\mathcal{K}_b}p_{bkc}= \rho_{bc^{'}}\sum_{k\in\mathcal{K}_b}p_{bkc^{'}},\notag\\
    &\,\,\,\,\,\,\,\,\,\,\,\,\,\,\,\,\,\,\,\,\,\,\,\,\,\,\,\,\,\,\,\,\,\,\,\,\,\,\,\,\,\,\,\,\,\,\,\,\,\,\,\,\,\,\,\,\forall c,c^{'}\in\mathcal{C}, c\neq c^{'},\label{con:pc6b}\\
    %&\rho_{bc}\sum_{k\in\mathcal{K}_b}p_{bkc}= P_b,\forall c\in\mathcal{C}, \label{con:pc6b}\\
    &\sum_{k\in\mathcal{K}_b}\alpha_{bkc}\leq \bar{K},\forall c\in\mathcal{C}, \label{con:al6b}\\
    &\sum_{c\in\mathcal{C}} \alpha_{bkc}=1, \forall k\in\mathcal{K}_b, \label{con:ak6b}\\
    &p_{bkc}\leq \hat{P}\alpha_{bkc},\forall c\in\mathcal{C},\forall k\in\mathcal{K}_b.\label{con:pa6b}
\end{align}
\end{subequations}
The decoding order indicators $\phi_{bklc}$ are determined based on $\mathbf{P}$ and $g_{bkc}$.
Thus variables $\phi_{bklc}$ are therefore fixed and constraints \eqref{con:phig1} and \eqref{con:phi1} are no longer needed in $\mathcal{P}_4(b)$.
By expressing $p_{bkc}$ by $R_{bkc}$, $\mathcal{P}_4(b)$ is reformulated as:
\begin{subequations}
\begin{align}
    \mathcal{P}_5(b):&\max_{R_{bkc},\alpha_{bkc},t_b}\,\,\,\,t_b \label{OPT14}\\
    \mbox{s.t.}\,\,\,\,
    &\eqref{con:al6b},\eqref{con:pa6b},\eqref{con:ak6b},\\
    &\sum_{k=1}^{K_b}\left(\dfrac{1}{g_{bkc}}-\dfrac{1}{g_{b(k-1)c}}\right) e^{\sum\limits_{j\geq k} \frac{R_{bjc}}{B_W}}-\dfrac{1}{g_{bK_bc}}\leq \dfrac{P_{b}}{\rho_{bc}},\notag\\
    &\,\,\,\,\,\,\,\,\,\,\,\,\,\,\,\,\,\,\,\,\,\,\,\,\,\,\,\,\,\,\,\,\,\,\,\,\,\,\,\,\,\,\,\,\,\,\,\,\,\,\,\,\,\,\,\,\,\,\,\,\,\,\,\,\,\,\,\,\,\,\,\,\,\,\,\,\,\,\,\,\,\,\,\,\,\,\,\,\,\,\,\,\,\,\,\,\forall c\in\mathcal{C}, \label{con:pc64}\\
    &t_bD_{bk}-R_{bk}\leq 0, \forall k\in\mathcal{K}_b, \label{con:rld14}
\end{align}
\end{subequations}
where the inequalities in \eqref{con:pc6b} are relaxed as the inequalities in \eqref{con:pc64} to convert the constraints into exponential cones.
Thus $\mathcal{P}_5(b)$ is identified as Mixed-Integer Exponential Conic Programming (MIECP) \cite{boyd2004convex}, whose optimum can be solved by branch and bound or outer approximation approach.

Similar to $f_b(\mathbf{P})=t_b^*$ in $\mathcal{P}_2(b)$, the optimal objective $\bar{t}_b^*$ in $\mathcal{P}_5(b)$ can be re-expressed by an inexplicit function of $\mathbf{P}$, say $\bar{f}_b(\mathbf{P})$.
Based on \textbf{Lemma 1} and \textbf{Lemma 2}, the objective function $f_b(\mathbf{P})$ in $\mathcal{P}_2(b)$ is a CUF. 
We then conclude that $\bar{f}_b(\mathbf{P})$ is also a CUF in \textbf{Corollary 2}. 

\textbf{Corollary 2.}
\textit{$\bar{f}_b(\mathbf{P})$ is a CUF.}

\begin{proof}
The properties of positivity and competitiveness follow analogously from \textbf{Lemma 1} and \textbf{Lemma 2}. 
Regarding the directional monotonicity, given $\zeta\mathbf{P}$ and $\mathbf{P}$ to $\mathcal{P}_5(b)$, we can obtain the optimal terminal-timeslot allocation $\boldsymbol{\alpha}_1^*$ and $\boldsymbol{\alpha}_2^*$, respectively, where $\boldsymbol{\alpha}_1^*$ and $\boldsymbol{\alpha}_2^*$ collect all $\alpha$-variables in beam $b$.
Note that the difference between $\mathcal{P}_5(b)$ and $\mathcal{P}_2(b)$ is that $\alpha_{bkc}$ is treated as fixed parameters in $\mathcal{P}_2(b)$, whereas $\alpha_{bkc}$ is to be optimized in $\mathcal{P}_5(b)$ as variables.
Thus, under the same $\boldsymbol{\alpha}_2^*$ in $\mathcal{P}_2(b)$, $f_b(\zeta \mathbf{P}) > f_b(\mathbf{P})$ can hold for $\zeta>1$ according to \textbf{Lemma 1} and \textbf{Lemma 2}.
Since $\boldsymbol{\alpha}_2^*$ is the optimal outcome of using $\mathbf{P}$ in $\mathcal{P}_5(b)$, then $\bar{f}_b(\mathbf{P})= f_b(\mathbf{P})$.
Compared with $f_b(\zeta\mathbf{P})$ and $\bar{f}_b(\zeta\mathbf{P})$,  $f_b(\zeta\mathbf{P})$ with a suboptimal $\boldsymbol{\alpha}_2^*$ is no higher than $\bar{f}_b(\zeta\mathbf{P})$ with its optimal $\boldsymbol{\alpha}_1^*$, thus $\bar{f}_b(\zeta\mathbf{P})>f_b(\zeta\mathbf{P})$, and $\bar{f}_b(\zeta\mathbf{P})>\bar{f}_b(\mathbf{P})$, then the conclusion.
\end{proof}

\begin{algorithm}[t]
\caption{JOPDT}
\label{alg:3}
\begin{algorithmic}[1]
\REQUIRE ~~\\
Initial beam power: $\mathbf{P}^{(0)}$,\\ Iteration index in the JOPD framework, $n=0$,\\ Iteration index: $\bar{n}=0$,\\ Maximum iteration: $\bar{N}_{\max}$,\\ Initial terminal-timeslot assignment: $\boldsymbol{\alpha}_b^{(0)}$, $\forall b\in\mathcal{B}$.
%\STATE \textit{Outer loop:}
\REPEAT 
\STATE Update precoding vectors $\mathbf{w}_{bc}$ based on $\boldsymbol{\alpha}_b^{(\bar{n})}$.
%\STATE \textit{Inner loop:}
\REPEAT
\STATE Update and sort $g_{bkc}$ with $\mathbf{P}^{(n)}$. 
\STATE Decide decoding orders ${\phi}_{bklc}$ based on the descending orders of $g_{bkc}$.
\STATE Solve $\mathcal{P}_5(b)$ and obtain $\bar{t}_b^{*(n)}$ with $\mathbf{P}^{(n)}$.
\STATE Update $\mathbf{P}$ by $P_b:=\frac{P_b^{(n)}}{\bar{t}_b^{*(n)}}$, $\forall b\in\mathcal{B}$.
\STATE Calculate $\epsilon=\max\left\{ \frac{P_b}{P_{b,\max}}, \forall b\in\mathcal{B}; \sum_{b\in\mathcal{B}}\frac{P_b}{P_{tot}}\right\}$.
\STATE Update $\mathbf{P}^{(n+1)}:=\frac{\mathbf{P}^{(n+1)}}{\epsilon}$. $n:=n+1$. 
\UNTIL{convergence}
\STATE $\bar{n}:=\bar{n}+1$. 
\STATE Update timeslot assignment $\boldsymbol{\alpha}_b^{(\bar{n})}$. 
\UNTIL{$\bar{n}>\bar{N}_{\max}$}
\STATE Calculate $p_{bkc}$ based on $\mathbf{P}^{(n)}$.
\ENSURE ~~\\
$\bar{t}_b^*$, $p_{bkc}$, $\boldsymbol{\alpha}_b$, $\mathbf{w}_{bc}$.
\end{algorithmic}
\end{algorithm}

Owing to the linearity, the constraints in the formulation of $\mathcal{P}_3$ are MCs.
With the properties of CUF and MC, the beam power allocation problem is an MCMU and can be tackled by the PF-based approach.
By solving $\mathcal{P}_5$ in line 6, a new terminal-timeslot assignment $\alpha_{b}^{(\bar{n})}$ is obtained (updated in line 12), and the optimal $\bar{t}^*_b(n)$ is achieved, which is used to update beam power in line 7.
The algorithm terminates when the number of iterations reaches $\bar{N}_{\max}$.

\section{Performance Evaluation}
\subsection{Parameter Settings}
\newcommand{\tabincell}[2]{\begin{tabular}{@{}#1@{}}#2\end{tabular}}  
\begin{table}[t]
\centering
\caption{Simulation Parameters}
  \begin{tabular}{|c|c|}
  %{@{}cccc@{}} \toprule
  \hline
  \textbf{Parameter} & \textbf{Value}\\\hline
  Frequency & 20 GHz (Ka band)\\\hline
  $B_W$ & 500 MHz\\\hline
  Satellite location & $13^{\circ}$ E\\\hline 
  Satellite height & 35,786 km\\\hline
  Satellite antenna gain & between 49.60 and 54.63 dBi\\\hline
  Receive antenna gain & 42.1 dBi\\\hline
  Output back off & 5dB\\\hline
  Channel model & LoS channel (path loss)\\\hline
  $\sigma^2$ & -126.47 dBW\\\hline
  $B$ & 4\\\hline
  $C$ & 5\\\hline
%  Freq. reuse  & 1 color\\\hline
  $P_{b,\max}$ & 120 W \cite{aravanis2015power}\\\hline
  $P_{tot}$   & 400 W \cite{aravanis2015power} \\\hline
  $\vert\mathcal{U}_b\vert$ & 70 \\\hline
  $\bar{K}$ & 2 \\\hline
  $D_{bk}$ & \tabincell{c}{Uniformly distributed\\between 0.5 and 3.5 Gbps} \\\hline
  %$N_{\max}$ & 15\\\hline
  $\bar{N}_{\max}$ & 5\\\hline
  \end{tabular}
  \label{tab:para}
\end{table}

\begin{figure}[t]
\centering
\includegraphics[scale=0.58]{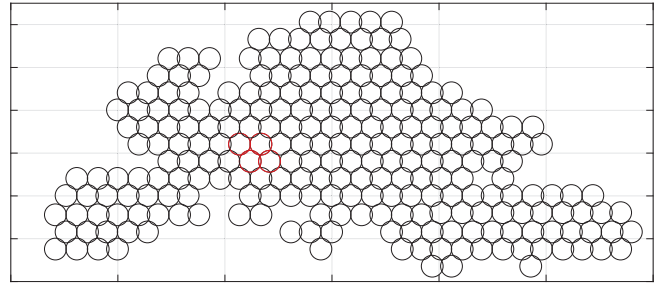}
\centering
\captionsetup{font={small}} 
\caption{Beam pattern covering Europe provided by ESA \cite{esa}. The figure shows an instance of four beams (highlighted in red color) served by an MPA.}
\label{fig:bp}
\end{figure}

We evaluate the performance of the proposed resource allocation approaches in a NOMA-enabled multi-beam satellite system.
The key parameters are summarized in Table \ref{tab:para}. 
The beam pattern provided by European Space Agency (ESA) \cite{esa} is illustrated in Fig. \ref{fig:bp}.
In the system, NOMA is applied in a small cluster of beams ($B=4$) which are served by an MPA.
Adjacent clusters occupy orthogonal frequencies such that the inter-cluster interference can be neglected.
Note that the variation of transmit antenna gain is related to the off-axis angle between the beam center and the terminal.
In NOMA, since the complexity of multi-user detection increases with the number of signals to be detected by the receiver \cite{perez2019non}, $\bar{K}=2$ is set in the simulation.
The results are averaged over 1000 instances. 
For each instance, one terminal is randomly selected from $\mathcal{U}_b$ and the other is paired via MaxCC for each timeslot.
%Besides, the considered cluster is randomly chosen for each instance.
Two NOMA-based schemes, i.e., JOPD+MaxCC with lower complexity and JOPDT with higher complexity, are compared to OMA and other benchmarks.

\subsection{Numerical Results}
\subsubsection{Convergence performance of JOPD}
\begin{figure}[t]
\centering
\subfigure[]{
\includegraphics[scale=0.35]{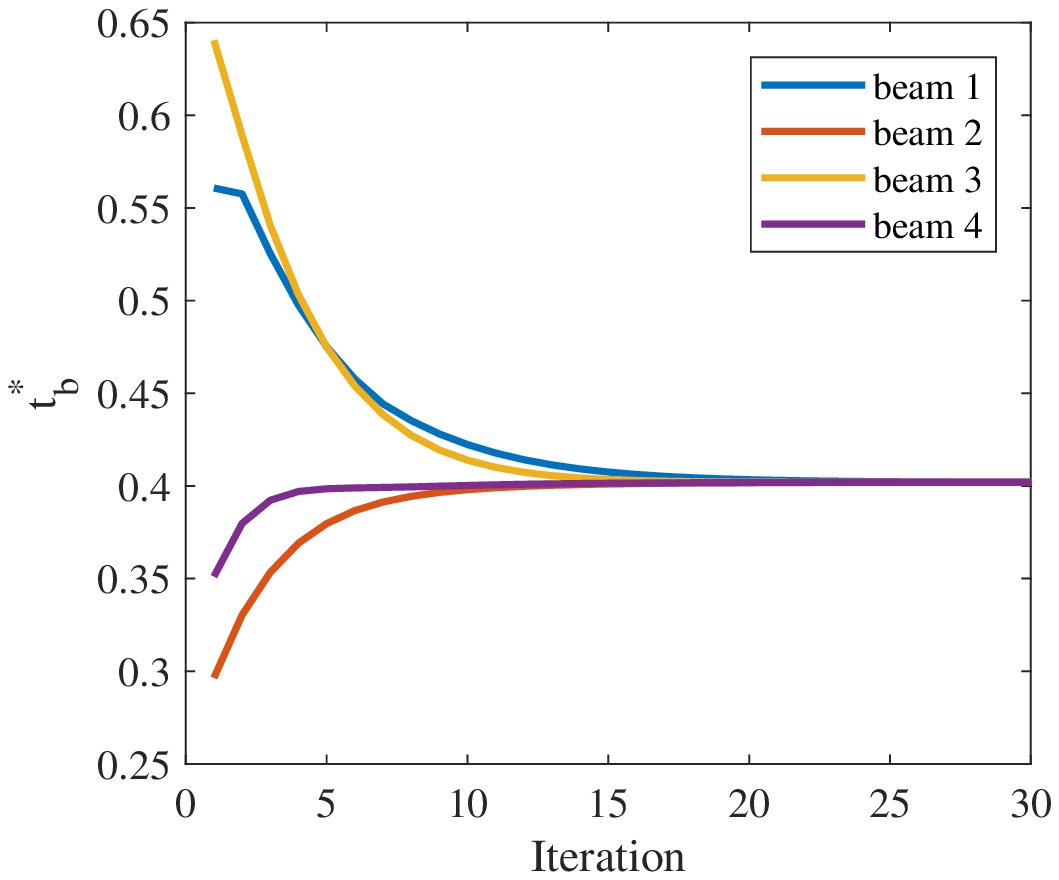}
\centering
\label{fig:t2}
}
\subfigure[]{
\includegraphics[scale=0.35]{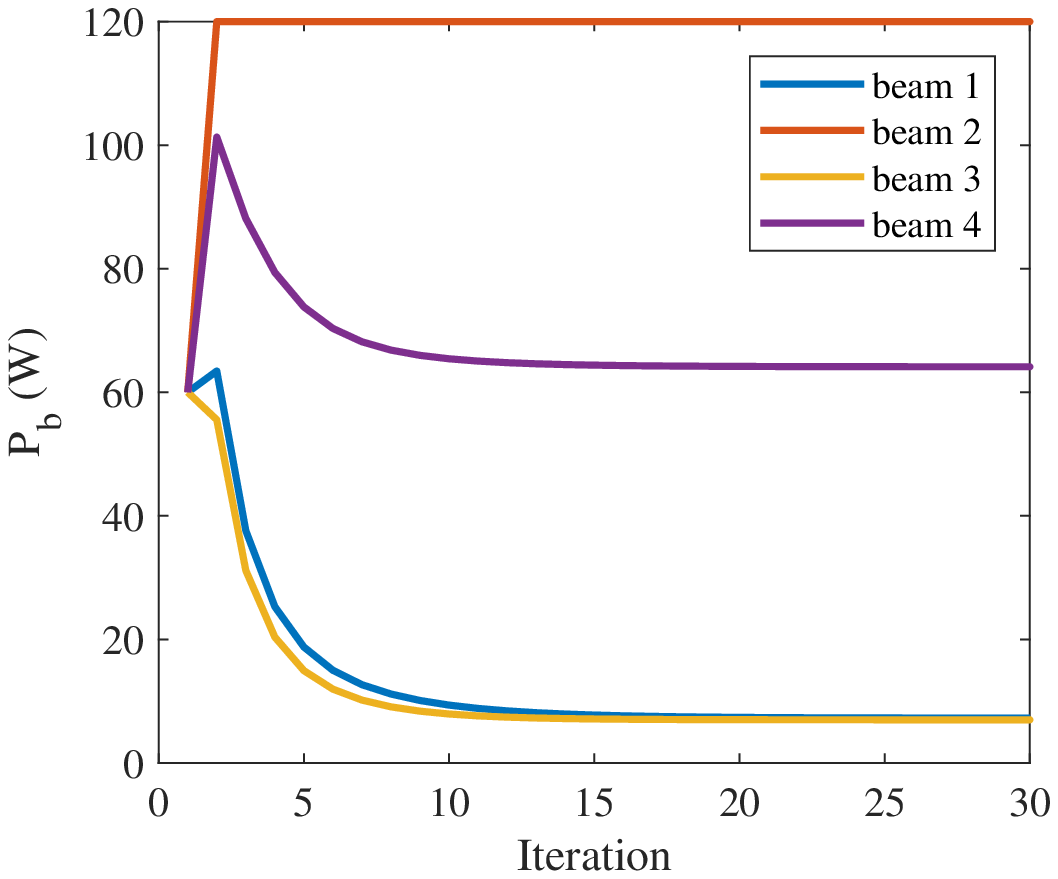}
\centering
\label{fig:p2}
}
\captionsetup{font={small}} 
\caption{Evolutions of $t_b^*$ and $P_b$ over iterations in JOPD.}
\label{fig:1}
\end{figure}

We first verify the convergence performance of JOPD.
Fig. \ref{fig:1} shows the evolutions of $t^*_b$ and $P_b$ over iterations.
From the figures, we observe that beam power is adjusted based on the values of $t^*_b$.
The power of the beams with smaller $t^*_b$ increases while the power of the other beams decreases in each iteration.
As it is proven in \textbf{Theorem 1}, JOPD converges, e.g., in Fig. \ref{fig:t2} within around 15 iterations.
Besides, the results verify the conclusion of \textbf{Lemma 2}, that is, the convergence of a CUF is not affected by the variation of decoding orders.

\subsubsection{Comparison of max-min OCTR between NOMA and OMA}
\begin{figure}[t]
\centering
\includegraphics[scale=0.68]{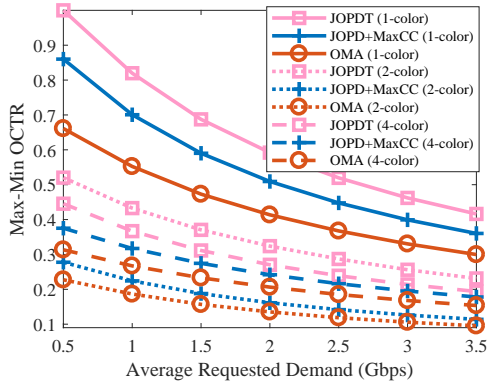}
\centering
\captionsetup{font={small}} 
\caption{Max-min OCTR with respect to traffic demand among JOPDT, JOPD+MaxCC, and OMA.}
\label{fig:4}
\end{figure}

Next, we compare the max-min OCTR performance among JOPDT, JOPD+MaxCC, and OMA in Fig. \ref{fig:4} to verify the superiority of the proposed NOMA-based schemes.
Different frequency-reuse patterns, i.e., 1-color, 2-color, and 4-color frequency-reuse patterns, are implemented.
In 1-color frequency-reuse pattern, the entire bandwidth is shared by all the spot beams.
2-color (or 4-color) pattern refers to the scenarios that the bandwidth is equally divided into 2 (or 4) portions, each of which is occupied by one of the 2 (or 4) adjacent beams. 
In OMA, the available frequency band is halved.
Each half of the band is occupied by one terminal at each timeslot.
Note that terminals are paired and scheduled to each timeslot by MaxCC in OMA.

In average, JOPD with MaxCC outperforms OMA with MaxCC in max-min OCTR by 24.0\%, 20.0\%, and 17.5\% under 1-color, 2-color, and 4-color pattern, respectively.
Particularly, with the implementation of 1-color pattern, the max-min OCTR in JOPD is 30.1\% higher than that in OMA when the average requested demand is 0.5 Gbps.
JOPD coordinated with precoding and MaxCC benefits from both reduced inter-beam and intra-beam interference compared to OMA.
Remark that in 2-color pattern, both JOPD+MaxCC and OMA are worse than other frequency-reuse patterns.
The reason is that compared to 2-color pattern, precoding is more effective in 1-color to mitigate strong inter-beam interference to a large extent, whereas 4-color pattern inherently receives much less inter-beam interference than that of 2-color pattern.
Besides, the OCTR performance of JOPD+MaxCC is compared with JOPDT.
By taking into account optimizing the terminal-timeslot assignment, JOPDT is able to improve the max-min fairness by approximately 16.2\%, 98.2\%, and 12.7\% under 1-color, 2-color, and 4-color reuse patterns, respectively.
The results validate the improvement of JOPDT over JOPD by iteratively updating the terminal-timeslot assignment. 

\begin{figure}[t]
\includegraphics[scale=0.68]{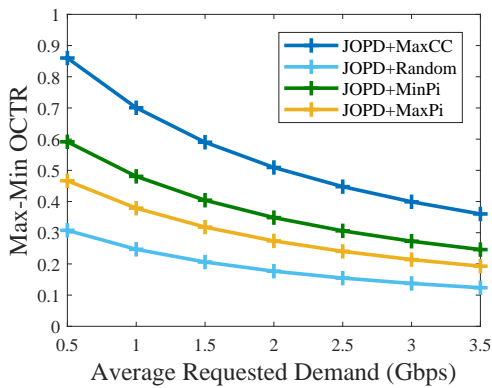}
\centering
\captionsetup{font={small}} 
\caption{Max-min OCTR with respect to traffic demand among different terminal-timeslot assignment approaches.}
\label{fig:5}
\end{figure}

\subsubsection{Comparison of max-min OCTR among different terminal-timeslot allocation approaches}
Different strategies of terminal-timeslot scheduling are compared in Fig. \ref{fig:5} in order to illustrate the advantages of MaxCC with NOMA in improving OCTR performance.
The basis of MaxCC is to allocate each timeslot to terminals with highest-correlation channels without considering the gap of $\Vert\mathbf{h}_{bk}\Vert$.
The benchmarks are listed as follows:
\begin{itemize}
\item MaxPi \cite{caus2016noma}: Allocate each timeslot to terminals with highly correlated channels and the largest gap of $\Vert\mathbf{h}_{bk}\Vert$,
\item MinPi \cite{caus2016noma}: Allocate each timeslot to terminals with highly correlated channels and the smallest gap of $\Vert\mathbf{h}_{bk}\Vert$,
\item Random: Allocate each timeslot to terminals randomly.
\end{itemize}
Note that in MaxPi and MinPi, terminals with the largest and smallest gain difference, respectively, are selected from those with correlation factor $\theta>0.9$.

From Fig. \ref{fig:5}, JOPD+MaxCC brings the largest gain compared to other benchmarks.
In MaxCC, the terminals with the highest channel correlation are selected.
Hence MaxCC can effectively reduce the inter-beam interference and exploit the synergy of NOMA with precoding.
Besides, the OCTR performance is sensitive to inter-beam interference.
The non-highest correlated channels in MinPi and MaxPi introduce a considerable amount of inter-beam interference and thus degrade the performance to a certain extent.

\begin{figure}[t]
\centering
\includegraphics[scale=0.58]{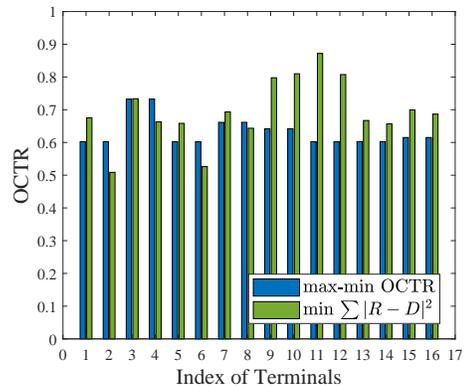}
\captionsetup{font={small}} 
\caption{The example of the distribution of OCTRs among terminals achieved by max-min OCTR and min $\sum_{b,k}\vert R_{bk}-D_{bk}\vert^2$ (16 terminals with 2.5 Gbps demand in average).}
\label{fig:useroctr}
\end{figure}

\subsubsection{Comparison between max-min OCTR metric and min $\sum_{b,k}\vert R_{bk}-D_{bk}\vert^2$ metric}
Lastly, we discuss the necessity of considering the max-min fairness of OCTR in multi-beam satellite systems.
Previous works, e.g., \cite{choi2005optimum}, focus on reducing the sum of the gap between offered capacity and requested traffic demand, i.e., min $\sum_{b,k}\vert R_{bk}-D_{bk}\vert ^2$.
Fig. \ref{fig:useroctr} presents the distribution of OCTRs among terminals achieved by JOPD+MaxCC, compared with NOMA to minimize $\sum_{b,k}\vert R_{bk}-D_{bk}\vert^2$.
The approach proposed in \cite{lei2019load} is adopted to solve the problem with the objective of min $\sum_{b,k}\vert R_{bk}-D_{bk}\vert ^2$.
We can observe that the max-min operator compromises the performance of high-capacity terminals, e.g., terminals 9 to 12, to compensate terminals with low OCTRs, e.g., terminals 2 and 6.
The average mismatch in max-min OCTR is relatively higher than min $\sum_{b,k}\vert R_{bk}-D_{bk}\vert^2$, but the terminals with worse channel conditions can get more resource, e.g., the minimum OCTR increases by 18.4\% than that in min $\sum_{b,k}\vert R_{bk}-D_{bk}\vert^2$ at the cost of losing 8.82\% of the average OCTR.

\section{Conclusion}
In this paper, we have introduced NOMA into multi-beam satellite systems to enable aggressive frequency reuse and enhance spectrum efficiency.
A max-min problem of jointly optimizing power, decoding orders, and terminal-timeslot assignment has been formulated to improve the worst OCTR among terminals.
We have proposed a PF-based algorithmic framework JOPD to jointly allocate power and decide decoding orders by fixing terminal-timeslot assignment with the guarantee of fast convergence.
Based on the framework of JOPD, a heuristic approach JOPDT has been developed to iteratively update the terminal-timeslot assignment and improve the overall OCTR performance.
The superiority of the proposed algorithms in max-min fairness over OMA has been demonstrated.
Besides, the numerical results have validated the applicability of the max-min OCTR metric in tackling practical issues in satellite scenarios.

%\bibliographystyle{IEEEtran}
%\bibliography{myref}

\begin{thebibliography}{99}

\bibitem{perez2019signal}
A.~I. P{\'e}rez-Neira, M.~{\'A}. V{\'a}zquez, M.~R.~B. {Shankar}, S.~Maleki,
  and S.~Chatzinotas, ``Signal processing for high-throughput satellites:
  {C}hallenges in new interference-limited scenarios,'' \emph{IEEE Signal
  Processing Magazine}, vol.~36, no.~4, pp. 112--131, 2019.

\bibitem{kodheli2020satellite}
O.~Kodheli \emph{et~al.}, ``Satellite communications in the new space era: A
  survey and future challenges,'' \emph{IEEE Communications Surveys \&
  Tutorials}.

\bibitem{simulator}
{SnT SIGCOM}, ``Satellite traffic emulator,''
  \url{https://wwwfr.uni.lu/snt/research/sigcom/sw_simulators}.

\bibitem{cocco2017radio}
G.~Cocco, T.~De~Cola, M.~Angelone, Z.~Katona, and S.~Erl, ``Radio resource
  management optimization of flexible satellite payloads for {DVB-S2}
  systems,'' \emph{IEEE Transactions on Broadcasting}, vol.~64, no.~2, pp.
  266--280, 2017.

\bibitem{lei2019learning}
L.~Lei, L.~You, Q.~He, T.~X. Vu, S.~Chatzinotas, D.~Yuan, and B.~Ottersten,
  ``Learning-assisted optimization for energy-efficient scheduling in
  deadline-aware {NOMA} systems,'' \emph{IEEE Transactions on Green
  Communications and Networking}, vol.~3, no.~3, pp. 615--627, 2019.

\bibitem{islam2017power}
S.~R. Islam, N.~Avazov, O.~A. Dobre, and K.-S. Kwak, ``Power-domain
  non-orthogonal multiple access ({NOMA}) in {5G} systems: Potentials and
  challenges,'' \emph{IEEE Communications Surveys \& Tutorials}, vol.~19,
  no.~2, pp. 721--742, 2017.

\bibitem{okamoto2016application}
E.~Okamoto and H.~Tsuji, ``Application of non-orthogonal multiple access scheme
  for satellite downlink in satellite/terrestrial integrated mobile
  communication system with dual satellites,'' \emph{IEICE Transactions on
  Communications}, vol.~99, no.~10, pp. 2146--2155, 2016.

\bibitem{caus2016noma}
M.~Caus, M.~{\'A}. V{\'a}zquez, and A.~I. P{\'e}rez-Neira, ``{NOMA} and
  interference limited satellite scenarios,'' in \emph{2016 50th Asilomar
  Conference on Signals, Systems and Computers}, 2016, pp. 497--501.

\bibitem{perez2019non}
A.~I. P{\'e}rez-Neira, M.~Caus, and M.~{\'A}. V{\'a}zquez, ``Non-orthogonal
  transmission techniques for multibeam satellite systems,'' \emph{IEEE
  Communications Magazine}, vol.~57, no.~12, pp. 58--63, 2019.

\bibitem{ugolini2019capacity}
A.~Ugolini, G.~Colavolpe, M.~Angelone, A.~Vanelli-Coralli, and A.~Ginesi,
  ``Capacity of interference exploitation schemes in multibeam satellite
  systems,'' \emph{IEEE Transactions on Aerospace and Electronic Systems},
  vol.~55, no.~6, pp. 3230--3245, 2019.

\bibitem{yan2019application}
X.~Yan, K.~An, T.~Liang, G.~Zheng, Z.~Ding, S.~Chatzinotas, and Y.~Liu, ``The
  application of power-domain non-orthogonal multiple access in satellite
  communication networks,'' \emph{IEEE Access}, vol.~7, pp. {63531--63539},
  2019.

\bibitem{zhu2017non}
X.~Zhu, C.~Jiang, L.~Kuang, N.~Ge, and J.~Lu, ``Non-orthogonal multiple access
  based integrated terrestrial-satellite networks,'' \emph{IEEE Journal on
  Selected Areas in Communications}, vol.~35, no.~10, pp. 2253--2267, 2017.

\bibitem{lin2019joint}
Z.~Lin, M.~Lin, J.-B. Wang, T.~de~Cola, and J.~Wang, ``Joint beamforming and
  power allocation for satellite-terrestrial integrated networks with
  non-orthogonal multiple access,'' \emph{IEEE Journal of Selected Topics in
  Signal Processing}, vol.~13, no.~3, pp. 657--670, 2019.

\bibitem{beigi2018interference}
N.~A.~K. Beigi and M.~R. Soleymani, ``Interference management using cooperative
  {NOMA} in multi-beam satellite systems,'' in \emph{2018 IEEE International
  Conference on Communications}, 2018, pp. 1--6.

\bibitem{yan2018outage}
X.~Yan, H.~Xiao, C.-X. Wang, and K.~An, ``Outage performance of {NOMA}-based
  hybrid satellite-terrestrial relay networks,'' \emph{IEEE Wireless
  Communications Letters}, vol.~7, no.~4, pp. 538--541, 2018.

\bibitem{liu2019qos}
X.~Liu, X.~Zhai, W.~Lu, and C.~Wu, ``{QoS}-guarantee resource allocation for
  multibeam satellite industrial internet of things with {NOMA},'' \emph{IEEE
  Transactions on Industrial Informatics}, 2019.

\bibitem{you2018resource}
L.~You, D.~Yuan, L.~Lei, S.~Sun, S.~Chatzinotas, and B.~Ottersten, ``Resource
  optimization with load coupling in multi-cell {NOMA},'' \emph{IEEE
  Transactions on Wireless Communications}, vol.~17, no.~7, pp. 4735--4749,
  2018.

\bibitem{liu2017joint}
Z.~Liu, L.~Lei, N.~Zhang, G.~Kang, and S.~Chatzinotas, ``Joint beamforming and
  power optimization with iterative user clustering for {MISO-NOMA} systems,''
  \emph{IEEE Access}, vol.~5, pp. 6872--6884, 2017.

\bibitem{chinnadurai2017novel}
S.~Chinnadurai, P.~Selvaprabhu, and M.~H. Lee, ``A novel joint user pairing and
  dynamic power allocation scheme in {MIMO-NOMA} system,'' in \emph{2017
  International Conference on Information and Communication Technology
  Convergence}, 2017, pp. 951--953.

\bibitem{ali2016non}
S.~Ali, E.~Hossain, and D.~I. Kim, ``Non-orthogonal multiple access ({NOMA})
  for downlink multiuser {MIMO} systems: User clustering, beamforming, and
  power allocation,'' \emph{IEEE Access}, vol.~5, pp. 565--577, 2016.

\bibitem{aravanis2015power}
A.~I. Aravanis, M.~R.~B. {Shankar}, P.-D. Arapoglou, G.~Danoy, P.~G. Cottis,
  and B.~Ottersten, ``Power allocation in multibeam satellite systems: A
  two-stage multi-objective optimization,'' \emph{IEEE Transactions on Wireless
  Communications}, vol.~14, no.~6, pp. 3171--3182, 2015.

\bibitem{wang2019fairness}
A.~Wang, L.~Lei, E.~Lagunas, A.~I. P{\'e}rez-Neira, S.~Chatzinotas, and
  B.~Ottersten, ``On fairness optimization for {NOMA}-enabled multi-beam
  satellite systems,'' in \emph{2019 IEEE 30th Annual International Symposium
  on Personal, Indoor and Mobile Radio Communications}, 2019, pp. 1--6.

\bibitem{christopoulos2015multicast}
D.~Christopoulos, S.~Chatzinotas, and B.~Ottersten, ``Multicast multigroup
  precoding and user scheduling for frame-based satellite communications,''
  \emph{IEEE Transactions on Wireless Communications}, vol.~14, no.~9, pp.
  4695--4707, 2015.

\bibitem{lei2019load}
L.~Lei, L.~You, Y.~Yang, D.~Yuan, S.~Chatzinotas, and B.~Ottersten, ``Load
  coupling and energy optimization in multi-cell and multi-carrier {NOMA}
  networks,'' \emph{IEEE Transactions on Vehicular Technology}, vol.~68,
  no.~11, pp. 11\,323--11\,337, 2019.

\bibitem{9014019}
M.~G. {Kibria}, E.~{Lagunas}, N.~{Maturo}, D.~{Spano}, H.~{Al-Hraishawi}, and
  S.~{Chatzinotas}, ``Carrier aggregation in multi-beam high throughput
  satellite systems,'' in \emph{2019 IEEE Global Communications Conference},
  2019, pp. 1--6.

\bibitem{destounis2011dynamic}
A.~Destounis and A.~D. Panagopoulos, ``Dynamic power allocation for broadband
  multi-beam satellite communication networks,'' \emph{IEEE Communications
  letters}, vol.~15, no.~4, pp. 380--382, 2011.

\bibitem{tan2015wireless}
C.~W. Tan \emph{et~al.}, ``Wireless network optimization by{ Perron-Frobenius}
  theory,'' \emph{Foundations and Trends{\textregistered} in Networking},
  vol.~9, no. 2-3, pp. 107--218, 2015.

\bibitem{huang2013joint}
Y.~Huang, C.~W. Tan, and B.~D. Rao, ``Joint beamforming and power control in
  coordinated multicell: Max-min duality, effective network and large system
  transition,'' \emph{IEEE Transactions on Wireless Communications}, vol.~12,
  no.~6, pp. 2730--2742, 2013.

\bibitem{zheng2017max}
L.~Zheng, D.~W. Cai, and C.~W. Tan, ``Max-min fairness rate control in wireless
  networks: Optimality and algorithms by {Perron-Frobenius} theory,''
  \emph{IEEE Transactions on Mobile Computing}, vol.~17, no.~1, pp. 127--140,
  2017.

\bibitem{sun2017optimal}
Y.~Sun, D.~W.~K. Ng, and R.~Schober, ``Optimal resource allocation for
  multicarrier {MISO-NOMA} systems,'' in \emph{2017 IEEE International
  Conference on Communications}, 2017, pp. 1--7.

\bibitem{boyd2004convex}
S.~Boyd and L.~Vandenberghe, \emph{Convex optimization}.\hskip 1em plus 0.5em
  minus 0.4em\relax Cambridge university press, 2004.

\bibitem{zheng2016wireless}
L.~Zheng, Y.-W.~P. Hong, C.~W. Tan, C.-L. Hsieh, and C.-H. Lee, ``Wireless
  max--min utility fairness with general monotonic constraints by
  {Perron--Frobenius} theory,'' \emph{IEEE Transactions on Information Theory},
  vol.~62, no.~12, pp. 7283--7298, 2016.

\bibitem{esa}
{ESA}, ``{SATellite Network of EXperts (SATNEX) IV},''
  \url{https://satnex4.org/}.

\bibitem{choi2005optimum}
J.~P. Choi and V.~W. Chan, ``Optimum power and beam allocation based on traffic
  demands and channel conditions over satellite downlinks,'' \emph{IEEE
  Transactions on Wireless Communications}, vol.~4, no.~6, pp. 2983--2993,
  2005.

\end{thebibliography}

\end{document}